\documentclass[lettersize,journal]{IEEEtran}
\usepackage{amsmath,amsfonts}
\usepackage{algorithmic}
\usepackage{algorithm}
\usepackage{array}
\usepackage[caption=false,font=normalsize,labelfont=sf,textfont=sf]{subfig}
\usepackage{textcomp}
\usepackage{stfloats}
\usepackage{url}
\usepackage{verbatim}
\usepackage{graphicx}
\usepackage{cite}
\usepackage{amsthm}
\usepackage{multirow}
\usepackage{makecell} 
\usepackage{soul}
\usepackage{tcolorbox}
\usepackage{colortbl}
\usepackage{xcolor}
\usepackage{pifont}
\usepackage{lipsum}
\usepackage{tikz}
\usepackage{fontawesome5}
\usepackage{longtable}
\usepackage{adjustbox}
\usepackage{booktabs}
\usepackage{threeparttable}

\definecolor{boxcolor}{HTML}{f3fbf2}
\definecolor{bestcolor}{HTML}{f3fbf2}
\definecolor{secondcolor}{HTML}{dff3f8}
\definecolor{3color}{HTML}{f7e8aa}
\definecolor{4color}{HTML}{ffdad2}
\definecolor{5color}{HTML}{d4e5ef}

\begin{document}

\title{Edge Intelligence with Spiking Neural Networks}


\author{Shuiguang Deng$^\dagger$,~\IEEEmembership{Senior Member,~IEEE},
        Di Yu$^\dagger$,
        Changze Lv,
        Xin Du,
        Linshan Jiang,
        Xiaofan Zhao,
        Wentao Tong,
        Xiaoqing Zheng,~\IEEEmembership{Member,~IEEE},
        Weijia Fang,
        Peng Zhao,
        Gang Pan,~\IEEEmembership{Senior Member,~IEEE},
        Schahram Dustdar,~\IEEEmembership{Fellow,~IEEE}, and Albert Y. Zomaya,~\IEEEmembership{Fellow,~IEEE}
    
    \thanks{· \textbf{Shuiguang Deng}, \textbf{Di Yu}, \textbf{Xiaofan Zhao}, \textbf{Wentao Tong}, and \textbf{Gang Pan} are with the College of Computer Science and Technology and with the State Key Laboratory of Brain-machine Intelligence, Zhejiang University, Hangzhou 310012, China (e-mail: dengsg@zju.edu.cn; yudi2023@zju.edu.cn;  xf.zhao@zju.edu.cn; toldzera@zju.edu.cn; gpan@zju.edu.cn).}
    \thanks{· \textbf{Xin Du} is with the School of Software Technology and with the State Key Laboratory of Brain-machine Intelligence, Zhejiang University, Hangzhou 310012, China. He is the Corresponding author of this research (e-mail: xindu@zju.edu.cn).}
    \thanks{· \textbf{Changze Lv} and \textbf{Xiaoqing Zheng} are with the School of Computer Science, Fudan University, Shanghai 200433, China (e-mail: czlv24@m.fudan.edu.cn; zhengxq@fudan.edu.cn).}
    \thanks{· \textbf{Linshan Jiang} is with the Institute of Data Science, National University of Singapore, 119077, Singapore (e-mail: linshan@nus.edu.sg).}
    \thanks{· \textbf{Weijia Fang} and \textbf{Zhao Peng} are with the First Affiliated Hospital, Zhejiang University School of Medicine, Hangzhou 310003, China (e-mail: weijiafang@zju.edu.cn; zhaop@zju.edu.cn)}
    \thanks{· \textbf{Schahram Dustdar} is with the Distributed Systems Group, Technische Universität Wien, 1040 Vienna, Austria and with ICREA Barcelona, Spain (e-mail: dustdar@dsg.tuwien.ac.at).}
    \thanks{· \textbf{Albert Y. Zomaya} is with the School of Computer Science, The University of Sydney, Sydney, NSW 2006, Australia (e-mail: albert.zomaya@sydney.edu.au).}
    \thanks{$^\dagger$ Both authors contributed equally to this research.}
}

\markboth{Journal of \LaTeX\ Class Files,~Vol.~14, No.~8, August~2021}%
{Shell \MakeLowercase{\textit{et al.}}: A Sample Article Using IEEEtran.cls for IEEE Journals}


\maketitle

\begin{abstract}
The convergence of artificial intelligence and edge computing has spurred growing interest in enabling intelligent services directly on resource-constrained devices. While traditional deep learning models require significant computational resources and centralized data management, the resulting latency, bandwidth consumption, and privacy concerns have exposed critical limitations in cloud-centric paradigms. Brain-inspired computing, particularly Spiking Neural Networks (SNNs), offers a promising alternative by emulating biological neuronal dynamics to achieve low-power, event-driven computation. This survey provides a comprehensive overview of Edge Intelligence based on SNNs (EdgeSNNs), examining their potential to address the challenges of on-device learning, inference, and security in edge scenarios. We present a systematic taxonomy of EdgeSNN foundations, encompassing neuron models, learning algorithms, and supporting hardware platforms. Three representative practical considerations of EdgeSNN are discussed in depth: on-device inference using lightweight SNN models, resource-aware training and updating under non-stationary data conditions, and secure and privacy-preserving issues.
Furthermore, we highlight the limitations of evaluating EdgeSNNs on conventional hardware and introduce a dual-track benchmarking strategy to support fair comparisons and hardware-aware optimization. Through this study, we aim to bridge the gap between brain-inspired learning and practical edge deployment, offering insights into current advancements, open challenges, and future research directions. To the best of our knowledge, this is the first dedicated and comprehensive survey on EdgeSNNs, providing an essential reference for researchers and practitioners working at the intersection of neuromorphic computing and edge intelligence.
\end{abstract}


\section{Introduction}

\IEEEPARstart{T}{he} rapid proliferation of artificial intelligence (AI), particularly machine learning (ML), has led to the development of increasingly complex models delivering state-of-the-art (SOTA) performance across diverse domains such as natural language processing (NLP) \cite{guo2024large}, computer vision (CV) \cite{zou2023object}, personalized recommendation \cite{liu2024benchmarking}, and time series forecasting \cite{liang2024foundation}. These AI-driven applications are now transforming various aspects of daily life.

Traditional AI applications rely heavily on the substantial computational and storage resources provided by the cloud. Massive volumes of data, often generated by Internet of Things (IoT) devices and other edge sources, are transmitted to centralized servers for processing and management. Users then access cloud-hosted data to obtain desired insights \cite{hua2023edge}. However, this cloud-centric approach faces critical challenges, including low throughput, high latency, bandwidth bottlenecks \cite{mohammadi2018deep}, and growing concerns about data privacy \cite{voigt2017eu}. To address these issues, edge computing has emerged as a promising paradigm by relocating AI services closer to data sources such as IoT devices and end users. By executing AI tasks on edge nodes, this approach helps reduce latency, alleviate bandwidth pressure, and improve system reliability and user experience.

Despite its advantages, conventional edge intelligence remains highly computation-intensive, imposing stringent demands on CPUs, GPUs, memory, and network bandwidth \cite{deng2020edge}. This resource dependency limits the deployment of advanced AI models on ubiquitous edge devices. Although devices such as smartphones are becoming more capable, they still struggle to support many deep learning models. For example, most voice assistants—including Apple Siri and Microsoft Cortana—depend on cloud services and become inoperable without network connectivity \cite{xu2021edge}. Furthermore, the escalating computational demands of AI models are outpacing the performance gains predicted by Moore’s Law \cite{frenkel2023bottom}, indicating an impending bottleneck for traditional computing architectures. As a result, there is an urgent need for novel, resource-efficient, and scalable solutions to support future edge intelligence.

\begin{figure*}[!t]
\centering
\includegraphics[width=\textwidth]{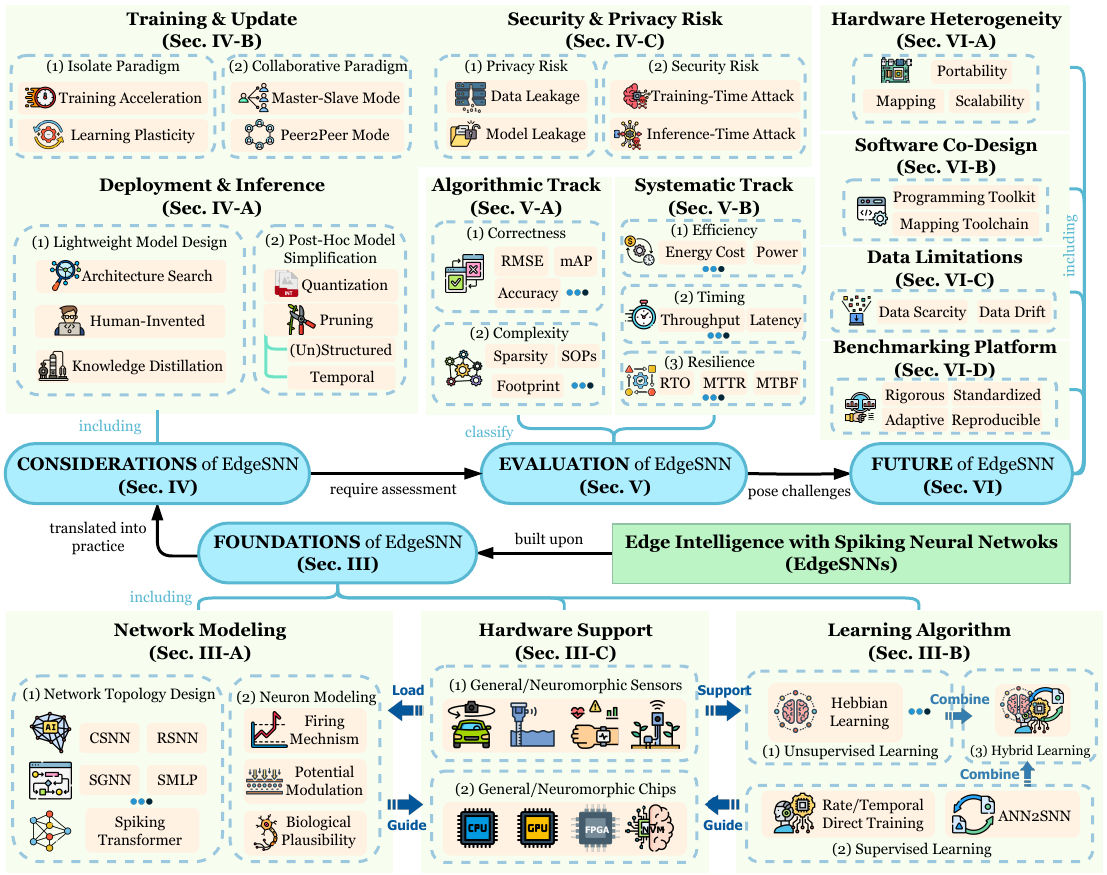}
\caption{The article structure of this survey is organized around four key aspects in the following order: (1) the foundational principles of EdgeSNNs, (2) ubiquitous practical considerations in building end-to-end EdgeSNN systems, (3) fair and rigorous evaluation methodologies tailored for EdgeSNNs, and (4) corresponding open challenges and future research directions.} 
\label{fig:survey-overview}
\end{figure*}

Brain-inspired computing (BIC) \cite{li2024brain} has emerged as a promising computing paradigm for addressing these challenges by incorporating neural principles and brain-inspired computational strategies into algorithms and systems \cite{schuman2022opportunities}.
In particular, brain-inspired algorithms leverage spiking neural networks (SNNs) \cite{maass1997networks} and neuroscience-derived mechanisms to enhance the learning plasticity and energy efficiency of computational models. Current algorithmic explorations are often conducted via simulated execution on conventional hardware platforms, such as CPUs and GPUs, to guide and refine the design of next-generation neuromorphic hardware \cite{yao2024spike}.
Brain-inspired systems \cite{frenkel2023bottom} play a pivotal role in advancing novel computing architectures, utilizing bio-inspired designs to enable scalable, energy-efficient, and real-time embodied computation. By integrating SNN-based algorithms with specialized neuromorphic hardware, these systems achieve significant improvements in energy efficiency, real-time performance, and resilience, thereby facilitating low-power intelligent applications at the edge \cite{yu2025eccsnn}.

To foster the future development of this emerging paradigm, which integrates edge computing with brain-inspired computing, referred to as Edge Intelligence based on Spiking Neural Networks (EdgeSNNs), it is essential first to identify the current advances in SNNs and assess their potential deployability in edge computing scenarios.
Existing research encompasses both single-neuron mechanisms at the cellular level and SNN architectures at the network level, along with their corresponding learning paradigms. Unlike artificial neural networks (ANNs), SNNs retain a consistent network topology while incorporating neuronal dynamics, wherein intrinsic biophysical processes govern spike generation. To address the variability in dynamical complexity, the Leaky Integrate-and-Fire (LIF) model \cite{gerstner2014neuronal} and its variants are commonly adopted as foundational neuron models to ensure computational efficiency.
Current learning algorithms for SNNs can be broadly categorized into unsupervised and supervised approaches \cite{rathi2018stdp,hao2023bridging,wu2019direct}, with the latter further divided into ANN-to-SNN conversion (ANN2SNN) methods and direct training techniques. A fundamental challenge in this area is the development of biologically plausible algorithms that can solve tasks difficult for contemporary AI models, particularly in resource-constrained edge environments. We argue that leveraging more biologically realistic neuron models holds significant promise for enhancing the adaptability and efficiency of neuromorphic computing systems. Nonetheless, their deployment in practical edge intelligence applications must be critically assessed across several key dimensions: task effectiveness, computational efficiency, and model trainability.

According to existing studies, we then categorize three practical real-world considerations for EdgeSNN applications:

\emph{\textbf{On-Device Deployment and Inference for EdgeSNNs}} focuses on deploying lightweight SNN-based models directly onto resource-constrained edge devices to enable low-latency, on-device inference while reducing energy consumption and dependence on cloud resources. This paradigm addresses critical issues such as response latency and communication overhead, which are particularly relevant in real-time edge applications. A central technical challenge lies in maintaining model accuracy while effectively reducing computational and memory demands. Existing solutions can be broadly categorized into lightweight model design \cite{liu2024lite, na2022autosnn} and post-hoc model simplification \cite{li2024towards,wei2025qp, qiu2025quantized}. These methods collectively aim to reduce the complexity of SNNs with negligible performance loss.

\emph{\textbf{Training and Updating EdgeSNNs}} involves on-device learning using locally available data, thereby reducing the privacy and security risks inherent in transmitting data to centralized servers. Nevertheless, the limited computational and memory resources of edge devices, coupled with potential shifts in data distributions \cite{ni2025alade}, present significant challenges to practical local training. To address these issues, two complementary strategies have been proposed: (1) enabling efficient on-device solo training with lower memory consumption and computational overhead \cite{anumasa2024enhancing}, thereby enhancing the feasibility of training and updating SNNs on resource-constrained edge platforms; and (2) facilitating collaborative learning among multiple edge devices \cite{ye2024asteroid} or with a central cloud server \cite{venkatesha2021federated}. In such collaborative settings, communication efficiency becomes a critical consideration, as both the frequency of updates and the associated transmission costs substantially influence the overall training effectiveness and model performance.

\emph{\textbf{Security and Privacy in EdgeSNNs}} aims at safeguarding on-device SNN-based models against potential malicious attacks and privacy breaches. On one hand, when local models engage in collaborative learning, there is a risk of sensitive information leakage. Although raw data are not explicitly shared, sophisticated analyses of model updates can potentially reveal private information. This risk is particularly acute in settings where the updates encapsulate distinctive patterns that reflect the underlying data distributions unique to individual edge devices or environments \cite{kim2022privatesnn, yadav2025differentially, kundu2024recent}. On the other hand, since the learning process relies heavily on local data collected at the edge, the system remains susceptible to adversarial manipulation \cite{bu2023rate}. An attacker could, for example, inject poisoned samples or adversarial inputs into the local dataset, thereby undermining the training process and leading to degraded or biased model behavior \cite{ding2022snn, liang2022toward, yao2024exploring}.

While these practical considerations lay the groundwork for EdgeSNNs, a fundamental challenge remains: how to establish comprehensive evaluation methodologies that ensure these models meet the diverse and stringent requirements of real-world edge applications.

Due to the nascent state of neuromorphic hardware, which has yet to converge on a commercially dominant platform, current evaluations of EdgeSNN performance are predominantly conducted on conventional edge devices such as Raspberry Pi and NVIDIA Jetson series\footnote{https://www.nvidia.com/en-us/autonomous-machines/embedded-systems/} \cite{ijcai2024p596,yu2024fedlec,yu2025eccsnn}. 
Although these devices effectively capture the heterogeneous resource constraints characteristic of edge platforms, evaluations of EdgeSNNs conducted on them remain limited and potentially biased, as such hardware is not explicitly designed to harness the full potential of BIC. In light of this, this study presents a dual-track evaluation scheme \cite{yik2025neurobench} to enable fair and rigorous assessment of EdgeSNN performance, which allows objective comparisons of novel SNN-based models across heterogeneous edge devices, guides future research toward hardware-aware optimization in real-world scenarios, and facilitates evidence-based commercialization strategies by quantifying the trade-offs between accuracy and resource constraints.

\begin{table}[!t]
\centering
\caption{Comparison with Existing Surveys.}
\label{tab:cmp-survey}
\resizebox{\columnwidth}{!}{%
\begin{tabular}{llm{1.2cm}m{5.7cm}}
\toprule
\rowcolor[HTML]{f3fbf2} 
\textbf{Ref.} & \textbf{Year} & \textbf{Venue} & {\raggedleft \textbf{Contribution (Pros \& Cons)}}  \\ 
\midrule \midrule
\rowcolor[HTML]{f3fbf2} 
\cite{schuman2022opportunities} & 2022 & Nature Computational Science & 
\textcolor{teal}{\faSmile[regular]} It offers a forward-looking perspective by envisioning large-scale future developments, including potential applications involving neuromorphic computing systems and the evolution of corresponding algorithmic support.   \par
\textcolor{red}{\faFrown[regular]} This study lacks detailed discussions of specific application concerns, comprehensive categorizations, and a systematic survey of the pertinent literature.
  \\
\midrule
\rowcolor[HTML]{f3fbf2} 
\cite{yamazaki2022spiking} & 2022 & MDPI Brain Sciences &  
\textcolor{teal}{\faSmile[regular]} The paper presents a detailed introduction to neuron models and training methodologies, laying a robust technical groundwork for subsequent developments. \par
\textcolor{red}{\faFrown[regular]} The study is limited in scope, addressing only computer vision and robotic control while neglecting other potential application domains.
  \\
\midrule
\rowcolor[HTML]{f3fbf2} 
\cite{rathi2023exploring} & 2023 & ACM Computing Survey &  
    \textcolor{teal}{
    \faSmile[regular]} The study offers a comprehensive and systematic examination of hardware design considerations in neuromorphic computing. \par
    \textcolor{red}{
    \faFrown[regular]} The paper inadequately covers applications of brain-inspired computing and provides limited in-depth analysis of network architectures from an algorithmic perspective.
  \\
\midrule
\rowcolor[HTML]{f3fbf2} 
\cite{li2024brain} & 2024 & Proceeding of IEEE &  
    \textcolor{teal}{\faSmile[regular]} The paper presents a comprehensive and systematic survey that covers key aspects, including software frameworks, datasets, and hardware platforms, in the BIC domain. \par 
    \textcolor{red}{\faFrown[regular]} The existing taxonomy of neuronal modeling is increasingly misaligned with current trends in SNN research, as it lacks comprehensive coverage of concrete application concerns, particularly those pertinent to edge computing contexts.
   \\
\midrule
\rowcolor[HTML]{f3fbf2} 
\cite{kudithipudi2025neuromorphic} & 2025 & Nature & 
    \textcolor{teal}{\faSmile[regular]} The work offers a forward-looking vision of large-scale neuromorphic systems, emphasizing the need for an integrated ecosystem encompassing software, hardware, and algorithms to support diverse future applications. \par 
    \textcolor{red}{\faFrown[regular]} It lacks a summary of historical contributions that have shaped the field’s current landscape.
 \\
\midrule
\rowcolor[HTML]{fff2e3} 
\textbf{Ours} & \textbf{2025} & \textbf{TBD} &  
    \textbf{Our paper presents one of the first comprehensive discussions on EdgeSNNs, covering algorithms, applications, and hardware. It also summarizes historical contributions and envisions a future on-device SNN ecosystem with categorized application concerns.}
   \\
 \bottomrule
\end{tabular}%
}
\end{table}

Despite encouraging advances, real-world EdgeSNN applications still present several open challenges spanning hardware, software, and algorithmic levels. 
Hardware heterogeneity, software immaturity, limited and dynamic data environments, and the lack of standardized benchmarking frameworks collectively hinder the development of efficient, adaptable, and scalable EdgeSNN solutions.
Addressing these challenges requires coordinated progress in co-design methodologies, software-hardware abstraction, adaptive learning under resource constraints, and standardized evaluation practices for edge intelligence.

\begin{figure*}[!t]
\centering
\includegraphics[width=\textwidth]{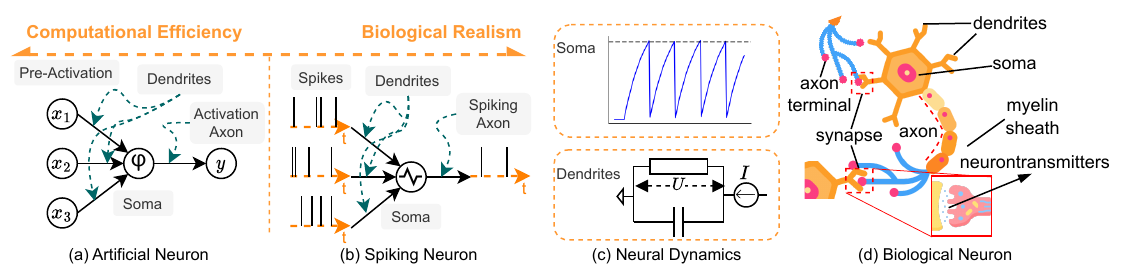}
\caption{Illustration of ANN and SNN neurons. 
(a) ANN neuron receives signals from the connected pre-neurons, conducts a nonlinear transformation $\phi(\cdot)$, and produces an output signal that multicasts to post-neurons.
(b) SNN neurons can be considered ANN neurons by substituting them with spiking neuronal dynamics.
(c) The soma dynamics are reflected in the membrane potential variations. The soma generates a spike when the membrane potential (\textcolor{blue}{blue line}) surpasses a threshold (\textcolor{gray}{gray dashed line}). The dendrite geometry can be divided into many compartments, each of which can be regarded as an RC circuit containing a capacitor to model the membrane and a resistor to model ion channels. (d) A typical biological neuron and synapse structure.} 
\label{fig:neurons}
\end{figure*}

In summary, this study provides a comprehensive and timely survey of methodologies developed for, or potentially applicable to, EdgeSNNs. 
The comparison of this survey with existing related surveys is shown in Table~\ref{tab:cmp-survey}. 
An accompanying GitHub repository\footnote{https://github.com/zju-bis/edgesnn-survey} is also provided, which compiles the papers included in this survey.
The main contributions of this survey are as follows:
\begin{itemize} 
    \item \textbf{Comprehensive Review}: To the best of our knowledge, this is the first survey dedicated to deploying practical SNN-based systems at the edge. It thoroughly reviews EdgeSNN technologies, encompassing fundamental methodologies, practical applications across various AI tasks, and deployment strategies in diverse edge scenarios.
    \item \textbf{Systematic Taxonomy}: A systematic taxonomy is introduced to categorize the core technological components of EdgeSNN development along with their corresponding application concerns. This taxonomy provides a structured framework for interpreting SOTA approaches and their underlying operational principles.
    \item \textbf{Challenges and Future Directions}: This study presents a comprehensive synthesis based on prior taxonomies and practical applications of EdgeSNNs, providing a critical analysis of prevailing challenges within current research. Through a systematic examination of existing limitations, we outline informed future research directions aimed at enhancing methodological rigor and addressing emerging trends that are poised to shape the evolution of this domain fundamentally.
\end{itemize}
Fig.~\ref{fig:survey-overview} depicts the paper structure.
The remainder of this survey is organized as follows. Section~\ref{sec:preliminary} introduces essential preliminary concepts related to SNNs.
Section~\ref{sec:foundation} provides an overview of recent advances in the foundation design of EdgeSNNs and discusses their suitability for edge scenarios.
Section~\ref{sec:apps} discusses diverse practical considerations for EdgeSNN implementation, encompassing deployment and inference, on-device training and updating, as well as privacy and security issues. Section~\ref{sec:eval} outlines a dual-track evaluation scheme designed to support efficient algorithmic and system-level development for EdgeSNNs. Section~\ref{sec:challenge} examines key challenges and outlines future research directions, concluding the survey in Section~\ref{sec:conclude}.

\section{Preliminaries} \label{sec:preliminary}

\subsection{Artificial Neural Networks vs. Spiking Neural Networks}

As depicted in Fig.~\ref{fig:neurons}, a neuron is the fundamental unit of a neural network that receives signals from connected pre-synaptic neurons, performs a nonlinear transformation, and generates an output signal that is transmitted to post-synaptic neurons. 
The connection between neurons is called a synapse, with the strength of the connection termed the weight $W$. The neuronal signal is called the activation $x$, and the nonlinear transformation is known as the activation function $\phi(\cdot)$.
The output $y_i$ of neuron $i$ can be expressed as:
\begin{align}
    y_i = \phi(\sum_j W_{ij}x_j + b_i) 
\end{align}
where $b_i$ is a bias. Then, we define the networks connected using the above neuron model as ANNs.

SNNs can be viewed as ANNs where conventional neurons or synaptic weights are replaced by spiking neuronal dynamics. The range of neuronal dynamics varies significantly, from simple first-order differential equations to more complex systems involving multiple differential equations. These dynamics sometimes extend beyond the soma to include dendritic processes. 
ANNs and SNNs share the same network topology, whereas the key distinction lies in the neurons of SNNs, which are governed by differential equations that model neuronal dynamics. Given their rich and dynamic properties, the spikes in SNNs can be closely linked to neural circuits, making them a natural fit for BIC \cite{li2024brain}, which is emerging as a promising energy-efficient alternative to traditional neural computing. Meanwhile, SNNs have already found widespread application in areas such as image and speech recognition \cite{wang2023complex,yao2021temporal}, object detection \cite{su2023deep}, autonomous driving \cite{zhu2024autonomous}, and other intelligence-driven real-world scenarios \cite{lv2023spiking,xiao24temporal}.

\subsection{Spike Coding Schemes for Real Inputs}
\label{sec:coding}

\begin{figure*}[!t]
\centering
\includegraphics[width=0.85\textwidth]{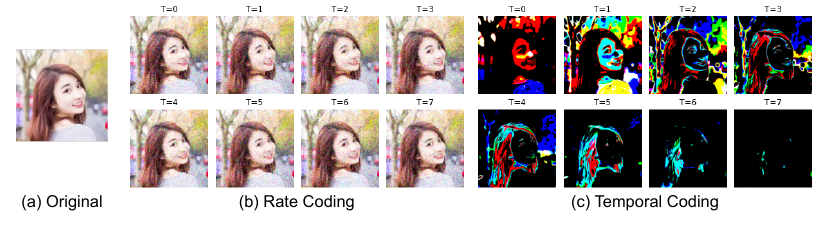}
\caption{Visualization of different coding schemes when the number of time steps is $8$. (a) The original RGB-based image. (b) Each input feature in the rate coding scheme indicates the probability that an event (spike) occurs at any given time step. (c) Temporal coding captures information about the precise firing time of neurons. High-intensity features fire before low-intensity ones.} 
\label{fig:code-type}
\end{figure*}

In practical EdgeSNN systems, although feeding non-spiking inputs into SNNs is common practice \cite{yao2023attention,shi2024spikingresformer,lv2024advancing}, encoding input data into spikes offers a more efficient representation. Spikes are inherently more suitable for storage and processing on neuromorphic hardware compared to high-precision continuous values.
Additionally, event-driven processing leads to sparsity, enabling faster processing and improved power efficiency by blocking unchanging input \cite{eshr2023train}. 
Apart from passing the input repeatedly at each time step like a static, unchanging video to SNNs (a.k.a. \textbf{direct/constant coding}), there are two standard data-to-spike coding methods \cite{auge2021survey} converting inputs into a spike sequence with a length equal to the number of time steps, as demonstrated in Fig.~\ref{fig:code-type}.

The first approach is \textbf{rate coding}, which utilizes each normalized input feature $x$ as the probability that a spike occurs at any given time step, resulting in a rate-coded value $r$ equal to 1, where $x\in [0,1]$.
This coding scheme represents information based on the average firing rate of a spiking neuron, whose process can be viewed as a Bernoulli trial $r\sim \mathrm{B}(1, x)$, a Poisson trial $r\sim \mathrm{Pois}(x)$ \cite{fang2023spikingjelly}, etc.
Although rate coding may not be optimal due to its higher spike frequency and increased power consumption \cite{olshausen2006other}, which neglects the temporal information carried by spikes, it exhibits high noise robustness in neuromorphic sensor scenarios. 

The other approach is \textbf{temporal coding}, which encodes information based on the precise timing of spikes.
The corresponding firing time occurs earlier when the stimulus intensity (input feature value $x$) increases. 
In particular, the logic of a temporal linear latency encoder (TTFS) can be denoted as:  
\begin{align}
    t_f(x)=T\times(1-x)
\end{align}
where $t_f$ is the specific spike-firing time within the permissible range of spike time window $T$ for a specific normalized feature $x\in [0, 1]$. 
This coding scheme requires only a single spike to convey information, effectively diminishing the number of spikes and significantly decreasing the energy requirements. Hence, it achieves higher sparsity and energy efficiency, making it a robust coding scheme that maximizes the energy efficiency of SNNs.

\section{Technical Foundations for EdgeSNNs} 
\label{sec:foundation}

Developing EdgeSNNs necessitates a reexamination of SNN foundations across network modeling, learning algorithms, and hardware design. Unlike conventional SNNs, which are optimized for accuracy or biological fidelity in unconstrained environments, EdgeSNNs must meet stringent constraints on energy, memory, and latency inherent to edge scenarios. This section reviews the foundational aspects of EdgeSNNs, highlighting their distinct design considerations in comparison to traditional SNNs.

\subsection{Modeling From Neuron to Network}

\subsubsection{Biological Neuron Modeling} \label{sec:neurons}
The primary function of a biological neuron is to integrate a multitude of synaptic inputs and convert them into a coherent stream of action potentials. As the fundamental unit of the nervous system, a neuron comprises four main components: dendrites, synapses, the soma, and the axon, as illustrated in Fig.~\ref{fig:neurons}(d). Dendrites receive input signals from other neurons and transmit them to the soma. When the resulting membrane potential exceeds a certain threshold, the soma generates a spike (action potential). This spike then propagates along the axon without attenuation and is transmitted to downstream neurons via synapses at the axon terminals.
Implementation cost is the primary concern when modeling a single neuron at the edge. This study evaluates the computational complexity of different neuron models from \textbf{spatial} and \textbf{temporal} perspectives.

In terms of spatial complexity, multi-compartment neuron models \cite{urbanczik2014learning, sacramento2018dendritic} incorporate neuronal morphology to enhance biological plausibility. However, this increased fidelity comes at the cost of substantially higher computational demands compared to single-compartment models, thereby limiting their practicality in edge intelligence applications.
In contrast, single-compartment models offer a more practical alternative by simplifying neuronal dynamics to focus exclusively on the soma, neglecting the structural and functional contributions of dendrites and axons. Among the various single-compartment models, the LIF model \cite{gerstner2014neuronal}, the Hodgkin-Huxley (HH) model \cite{hodgkin1952quantitative}, and the Izhikevich model \cite{izhikevich2003simple} are among the most widely adopted.

From the perspective of temporal dynamics, neuronal models can be broadly classified into two categories based on whether they incorporate temporal behavior. Artificial neurons (ANs), for instance, generate instantaneous outputs without modeling temporal evolution \cite{poirazi2001impact}. In contrast, spiking neuron models simulate the temporal dynamics of the membrane potential within the soma, producing spike trains over a defined time window. Among them, the LIF model is the most widely adopted in SNNs \cite{maass1997networks}. It captures key neuronal behaviors such as membrane potential integration, passive leakage, and threshold-triggered firing. The following equations typically describe the model’s dynamics:
\begin{align} \label{eq:update-mem}
     U[t] & = \tau\cdot H[t-1] + I[t] \\ 
     \label{eq:heaviside}
     O[t] &= \Theta(U[t] - \bar{V}) \\ 
     \label{eq:fire}
     H[t] &= U[t]\cdot (1-O[t]) + V_r\cdot O[t]
\end{align}
where $\tau, \bar{V}$, and $V_r$ denote the membrane potential decay factor (a.k.a. membrane time constant), threshold, and reset potential. $I[t]$ is the spatial input at time step $t$, $U[t]$ and $H[t]$ are the corresponding membrane potential before and after firing. Additionally, $\Theta(\cdot)$ is the Heaviside step function determining whether a spike is generated.
The LIF model enables lower power consumption by encoding signals as binary events. 

\begin{figure}[!t]
\centering
\includegraphics[width=\columnwidth]{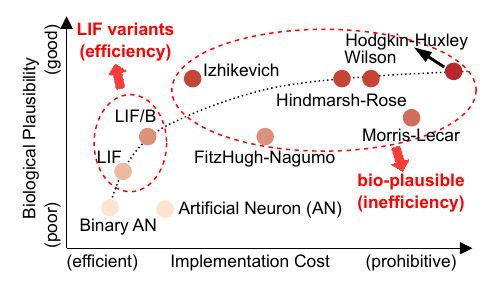}
\caption{A comparison of spiking neuron models \textit{w.r.t.} implementation cost and biological plausibility (adapted from \cite{izhikevich2004model}).}
\label{fig:history}
\end{figure}

Fig.~\ref{fig:history} compares the biological features of neuronal models and their energy cost for implementation.
In particular, models with higher bio-plausibility will consequently consume more energy. 
The LIF model incorporates more biological features than a typical artificial neuron and offers lower energy consumption by replacing the costly multiplication-and-accumulation (MAC) operation with the more efficient accumulation (AC) operation. 
However, it should be noted that current von Neumann architectures are better suited for performing dense matrix-vector multiplications \cite{owens2008gpu}, which are commonly used in ANs for AI tasks. 
Notably, a binary AN is the simplest spiking neuron without temporal dynamics and consumes the least energy cost. 
For applications of EdgeSNNs, related research may focus on the lower left corner circled in Fig.~\ref{fig:history} to boost computational efficiency and reduce power consumption.
Consequently, existing studies focus on improving the LIF model, resulting in various derived variants that enhance both bio-plausibility and effectiveness while maintaining the original computational efficiency. We summarize various recently proposed LIF variants that have the potential to be applied in EdgeSNNs in Table~\ref{tab:sum-neurons}. Based on these variants, we identify three broad extension directions for the basic LIF model.

One key direction involves \textbf{modifying the neuronal spike-firing mechanism} to enhance both functional expressiveness and computational efficiency. From a functional standpoint, since information in SNNs is transmitted through binarized spikes, conventional LIF-based networks are often regarded as inadequate for addressing complex machine learning tasks due to their limited representational capacity.
Therefore, some studies \cite{hammouamri2024learning,qiu2025quantized,yao2022glif,bal2024p} attempt to enhance the feature extraction capability of SNNs by introducing more diverse feature encoding schemes. For example, RSN \cite{chen2021deep} replaces binary spikes with ion counts, while T-LIF \cite{guo2024ternary} and ST-BIF \cite{you2025vistream} introduce ternary spikes, encoding information with \{-1, 0, 1\}. These approaches enhance the information encoding capacity of SNNs while retaining their compatibility with multiplication-addition operations. 
From an efficiency perspective, the sequential charge-fire-reset computation paradigm of the LIF model often results in slow simulation speeds. To address this issue, several studies \cite{taylor2023addressing, hu2024high, huang2024prf} have proposed optimizations aimed at reducing computational overhead. For example, the PSN model \cite{fang2023parallel} eliminates the reset operation to facilitate parallel processing.
However, the design of such variants remains a contentious issue. While they prioritize task-specific functionality and computational efficiency, they often neglect the fundamental principles of biological neuron modeling. This misalignment may result in poor compatibility with existing neuromorphic hardware \cite{yao2024spike, yang2024vision}, ultimately undermining the energy-efficiency advantages that SNNs are intended to deliver.

\begin{table}[!t]
\centering
\caption{Summary of LIF variants for SNN.}
\label{tab:sum-neurons}
\resizebox{\columnwidth}{!}{%
\begin{tabular}{lllr}
\toprule
\textbf{Neuron Model} & \textbf{Year} & \textbf{Venue} & \textbf{Proposed Tasks} \\ 
\midrule \midrule 
\rowcolor[HTML]{f3fbf2}
\multicolumn{4}{c}{\textbf{Firing Mechanism}} \\
\midrule
RSN \cite{chen2021deep} & 2021 & AAAI & Static Image Classification \\ 
GLIF \cite{yao2022glif} & 2022 & NeurIPs & Static \& Event-based Image Classification \\
ALIF \cite{taylor2023addressing} & 2023 & NeurIPs & Event-based Image Classification \\ 
PSN \cite{fang2023parallel} & 2023 & NeurIPs & Static \& Event-based Image Classification \\ 
T-LIF \cite{guo2024ternary} & 2024 & AAAI & Static \& Event-based Image Classification \\
T-RevSNN \cite{hu2024high} & 2024 & ICML & Static \& Event-based Image Classification \\ 
I-LIF \cite{lei2024spike2former} & 2025 & AAAI & Image Segmentation \\
ST-BIF \cite{you2025vistream} & 2025 & CVPR & Object Detection \& Tracking \& Segmentation \\
IE-LIF \cite{qiu2025quantized} & 2025 & ICLR & Static \& Event-based Image Classification \\
P-SpikeSSM \cite{bal2024p} & 2025 & ICLR & Image Classification \& Speech Recognition \\
\midrule
\rowcolor[HTML]{f3fbf2}
\multicolumn{4}{c}{\textbf{Potential Modulation}} \\
\midrule
PLIF \cite{fang2021incorporating} & 2021 & ICCV & Static \& Event-based Image Classification \\ 
SRIF \cite{guo2022reducing} & 2022 & ECCV & Static \& Event-based Image Classification \\
LTMD \cite{wang2022ltmd} & 2022 & NeurIPs & Static \& Event-based Image Classification \\ 
BDETT \cite{ding2022biologically} & 2022 & NeurIPs & Robotic Control \& Static Image Classification  \\ 
DTA-TTFS \cite{wei2023temporal} & 2023 & ICCV & Static Image Classification \\ 
CLIF \cite{huang2024clif} & 2024 & ICML & Static \& Event-based Image Classification \\
ABN \cite{kachole2024asynchronous} & 2024 & ECCV & Event-based Image Classification \& Segmentation \\
DA-LIF \cite{zhang2025lif} & 2025 & ICASSP & Static \& Event-based Image Classification \\
Smooth LIF \cite{ding2025rethinking} & 2025 & ICLR & Image Classification \& Speech Recognition \\
\midrule
\rowcolor[HTML]{f3fbf2}  
\multicolumn{4}{c}{\textbf{Biological Plausibility}} \\
\midrule
TC-LIF \cite{zhang2024tc} & 2024 & AAAI & Speech Recognition  \\ 
IHC-LIF \cite{song2024spiking} & 2024 & ICASSP & Speech Recognition \\
LM-H \cite{hao2023progressive} & 2024 & ICLR & Static Image Classification \\ 
STC-LIF \cite{wang2024autaptic} & 2024 & ICML & Traffic Flow Prediction \& Video Recognition \\
DH-LIF \cite{zheng2024temporal} & 2024 & NC & Speech Recognition \\
TS-LIF \cite{shibo2025tslif} & 2025 & ICLR & Time Series Prediction \\
\bottomrule
\end{tabular}%
}
\end{table}

Some studies also seek to enhance the representational capacity of LIF neurons by \textbf{modulating membrane potential settings} and then affecting neuronal spiking frequencies, primarily focusing on two core rules: 

(1) Adjusting the critical membrane potential dynamically. To enhance the representational capacity of SNNs, numerous studies have investigated the use of learnable threshold potentials, allowing the network to adjust its firing criteria during training for improved performance.
LTMD \cite{wang2022ltmd} allows the threshold potential to be learned across layers during training and presents a hyperbolic tangent relation to stabilize neuronal activity, thereby enhancing performance across various tasks.
ABN \cite{kachole2024asynchronous} offers network-independent firing threshold control based on historical activities of the membrane to maintain spiking stability.
BDETT \cite{ding2022biologically} incorporates layer-wise statistical cues to regulate the threshold potentials of neurons across different layers, thereby improving the generalization ability of SNNs.
Inspired by the temporal coding scheme, DTA-TTFS \cite{wei2023temporal} introduces a dynamic threshold potential setting that varies across network layers and time steps, effectively mitigating the over-sparsity caused by asynchronous transmission and alleviating the resource-intensive computations required for identifying presynaptic and postsynaptic neurons during training. 
In contrast, SRIF \cite{guo2022reducing} introduces a soft reset mechanism that drives the membrane potential toward a dynamic reset potential, ensuring the firing stability of spiking neurons and mitigating information loss.

(2) Refining the subthreshold dynamics of membrane potential. This direction is often related to the membrane time constant $\tau$ in Equation~(\ref{eq:update-mem}), which governs the dynamic transition rate of the membrane potential from the reset potential to the action potential threshold. 
For instance, PLIF \cite{fang2021incorporating} treats the time constant as a learnable parameter during the backpropagation-based learning phase, enabling better representation of neuronal heterogeneity and enhancing the expressiveness of SNNs.
DA-LIF \cite{zhang2025lif} introduces learnable spatial and temporal decay factors to enable more precise regulation of membrane permeability across SNN layers, enhancing performance while maintaining computational efficiency.
In \cite{ding2025rethinking}, a layer-shared smoothing coefficient is employed to weight the membrane state of the previous timestep, adaptively reducing membrane potential differences and enhancing output consistency across time steps, thereby facilitating the propagation of forward information.
Instead of modifying the decay factor, some works \cite{zhang2025enhancing,jiang2023klif} also explore modifying the membrane potential dynamics to alter the intrinsic spiking frequency patterns of neurons, e.g., CLIF \cite{huang2024clif} introduce extra paths for the complementary membrane potential to capture more temporal information and address the gradient errors vanishing problem in backpropagation.
Although LIF variants based on membrane potential modulation can enhance the overall representational capacity of SNNs by stabilizing spiking dynamics, these modifications also introduce additional computational overhead for model training or inference, which may limit their deployment in resource-restricted edge scenarios.

\begin{figure}[!t]
\centering
\includegraphics[width=0.8\columnwidth]{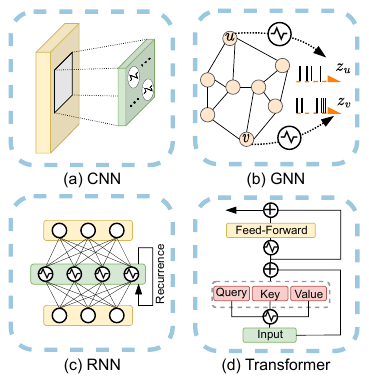}
\caption{Representative neural network topologies widely adopted in modern EdgeSNN systems. (a) \textbf{CNN} for spatial feature extraction, (b) \textbf{GNN} for learning over graph structures, (c) \textbf{RNN} for modeling sequential information, and (d) \textbf{Transformer} for capturing global dependencies via self-attention.} 
\label{fig:networks}
\end{figure}

The third extension direction focuses on \textbf{enhancing the biological plausibility} of LIF neurons by incorporating additional compartments, such as dendrites, to simulate a broader range of biological characteristics. This would thereby expand the representation space and improve neuronal heterogeneity.
TC-LIF \cite{zhang2024tc} and LM-H \cite{hao2023progressive} introduce independent soma and dendrite compartments along with corresponding learnable parameters, significantly enhancing the representational capacity of long-term dependencies. A similar design is further extended in IHC-LIF \cite{song2024spiking} to extract multi-scale temporal information efficiently, achieving optimal performance in speech signal processing tasks.
STC-LIF \cite{wang2024autaptic} integrates axon-dendrite and axon-soma circuits into LIF neurons, improving their ability to model spatiotemporal information in traffic flow.
TS-LIF \cite{shibo2025tslif} and DH-LIF \cite{zheng2024temporal} extend soma modeling of LIF by incorporating dual or multiple dendrite compartments to capture temporal features at different frequencies, thereby enhancing performance in time-driven sequential tasks. 
In summary, LIF variants developed along this direction are primarily tailored for specific purposes. However, their generalization ability across diverse domains remains an open question, necessitating extensive experimental validation to evaluate their effectiveness.

\subsubsection{Network Topology Designing}

Integrating the aforementioned neuron models into neural networks is critical for enabling advanced functionalities in edge environments \cite{ijcai2024p596}. As discussed in Section~\ref{sec:neurons}, the LIF model and its variants are widely adopted in contemporary SNNs due to their simplified morphology and event-driven nature, which contribute to energy efficiency. Fig.~\ref{fig:networks} presents four representative network architectures—Convolutional Neural Networks (CNNs), Graph Neural Networks (GNNs), Recurrent Neural Networks (RNNs), and Transformers—alongside others such as Multilayer Perceptrons (MLPs). While these networks traditionally employ artificial neurons, they can be adapted into SNNs by incorporating spatiotemporal dynamics at the neuron level. Ultimately, the choice of neuron model within a given architecture depends on the research objective and the demands of the target application.

CNNs, characterized by their hierarchical architecture and localized receptive fields, mirror the structure of the biological visual system, making them fundamental to contemporary computer vision research \cite{lecun2015deep}.
When integrated with spiking neurons, the resulting Convolutional Spiking Neural Networks (CSNNs) combine the representational power of CNNs with the event-driven, energy-efficient processing of neuromorphic systems.
CSNNs have been extensively applied across a wide range of vision tasks, including image classification \cite{yao2021temporal,fang2021deep,ren2023spiking,qiu2024gated,hu2024advancing}, object detection \cite{su2023deep,fan2024sfod,luo2024integer,li2025brain}, semantic segmentation \cite{patel2021spiking,meftah2010segmentation,zhang2024accurate}, and motion detection \cite{lutes2024convolutional,liu2021event,berlin2022spiking}, etc., as summarized in Table~\ref{tab:cnn-app}. 
These models demonstrate competitive performance while offering significantly lower power consumption compared to conventional deep learning methods. Furthermore, the intrinsic temporal coding capabilities of CSNNs enhance their ability to process dynamic visual stimuli, making them particularly well-suited for real-time and energy-constrained vision applications \cite{trivedi2025intelligent}.

\begin{table}[!t]
\centering
\caption{The summary of EdgeSNN applications with CNN topology.}
\label{tab:cnn-app}
\resizebox{\columnwidth}{!}{%
\begin{tabular}{lllr}
\toprule
\rowcolor[HTML]{f3fbf2}
\textbf{Model} & \textbf{Year} & \textbf{Venue} & \textbf{Proposed Tasks} \\
\midrule \midrule
SpikingVGG16 \cite{zhou2021temporal} & 2021 & AAAI & Static Image Classification \\
TA-SNN \cite{yao2021temporal} & 2021 & ICCV & Event-based Image Classification \\  
MotionSNN \cite{liu2021event} & 2021 & IJCAI & Human Action Recognition \\ 
SEW-ResNet \cite{fang2021deep} & 2021 & NeurIPs & Static \& Event-based Image Classification \\  
SparseSCNN \cite{yin2021energy} & 2021 & KDD & Static \& Event-based Image Classification \\ 
EMS-YOLO \cite{su2023deep} & 2023 & ICCV & Object Detection \\ 
TextCSNN \cite{lv2023spiking} & 2023 & ICLR & Natural Language Processing \\ 
Spiking PointNet \cite{ren2023spiking} & 2023 & NeurIPs &  3D object classification \\  
P2SResLNet \cite{wu2024point} & 2024 & AAAI & 3D object classification \\ 
GAC-SNN \cite{qiu2024gated} & 2024 & AAAI & Static Image Classification \\  
SFOD \cite{fan2024sfod} & 2024 & CVPR & Object Detection \\
SpikeYOLO \cite{luo2024integer} & 2024 & ECCV & Object Detection \\ 
ESDNet \cite{song2024learning} & 2024 & IJCAI & Image Deraining \\ 
Spike-TCN \cite{lv2024efficient} & 2024 & ICML & Time Series Prediction \\
SAD \cite{zhu2025autonomous} & 2024 & NeurIPs & Autonomous Driving \\ 
SpikingYOLOX \cite{miao2025spikingyolox} & 2025 & AAAI & Object Detection \\
E-3DSNN \cite{qiu2025efficient} & 2025 & AAAI & 3D Classification \& Detection \& Segmentation \\
CREST \cite{mao2025crest} & 2025 & AAAI & Object Detection \\
MSF \cite{qian2025ucf} & 2025 & AAAI & Video Anomaly Detection \\
\bottomrule
\end{tabular}%
}
\end{table}

GNNs have shown strong capabilities in learning structured representations from graph-structured data \cite{wu2022graph}. However, conventional GNNs typically involve dense matrix operations, resulting in high computational and memory costs. To mitigate these limitations, sparsification techniques—such as ReLU-based edge filtering—are often employed to reduce redundant connections while maintaining model expressiveness \cite{jin2024survey}. Conversely, SNNs naturally exhibit sparse, binary communication, positioning them as an energy-efficient alternative for graph-based learning in low-power, mobile, and hardware-constrained environments \cite{li2024graph,snyder2024transductive}. As outlined in Table~\ref{tab:gnn-app}, this biologically plausible computation paradigm is particularly well-aligned with the inherent sparsity of real-world graphs, enabling efficient solutions for tasks such as node classification \cite{chen2025signn,ijcai2021p0441}, relational reasoning \cite{xiao2024temporal}, and structural modeling \cite{sunspiking,ijcai2022p0338}.
Recent progress in Spiking GNNs (SGNNs) also highlights their promise in dynamic and temporal graph learning scenarios \cite{li2023scaling,zhao2024dynamic,yin2024dynamic}, where the event-driven nature of SNNs enables effective modeling of evolving graph structures. Despite these advances, scaling SGNNs to large graphs remains a significant challenge, marking an important direction for future research at the intersection of neuromorphic computing and graph representation learning \cite{wu2024vt}.

\begin{table}[!t]
\centering
\caption{The summary of EdgeSNN applications with GNN topology.}
\label{tab:gnn-app}
\resizebox{\columnwidth}{!}{%
\begin{tabular}{lllr}
\toprule
\rowcolor[HTML]{f3fbf2}  
\textbf{Model} & \textbf{Year} & \textbf{Venue} & \textbf{Proposed Tasks} \\
\midrule \midrule
GC/GA-SNN \cite{ijcai2021p0441} & 2021 & IJCAI & Node Classification \\
SpikingGCN \cite{ijcai2022p0338} & 2022 & IJCAI & Node Classification \& Recommendation \\ 
SpikeNet \cite{li2023scaling} & 2023 & AAAI & Node Classification \\   
DRSGNN \cite{zhao2024dynamic} & 2024 & AAAI & Node Classification \\ 
Dy-SIGN \cite{yin2024dynamic} & 2024 & AAAI & Node Classification \\ 
SpikeGCL \cite{li2024graph} & 2024 & ICLR & Node Classification \\
GRSNN \cite{xiao2024temporal} & 2024 & ICML & Relation/Link Prediction \\
MSG \cite{sunspiking} & 2024 & NeurIPs & Node Classification  \\
ASGCN \cite{zeng2025leveraging} & 2025 & AAAI & Event-based Image Classification \\
MSF \cite{qian2025ucf} & 2025 & AAAI & Video Anomaly Detection \\
NI-SGCN \cite{li2025noise} & 2025 & AAAI & 3D Point Cloud Denoising \\
\bottomrule
\end{tabular}%
}
\end{table}

Recurrent Spiking Neural Networks (RSNNs) have emerged as a potent computational framework for modeling complex dynamical systems, offering substantial potential across various sequential processing tasks. Their distinctive features—temporal coding and event-driven computation—make them particularly well-suited for neuromorphic systems that require efficient and bio-inspired processing of real-world sequential data. In robotics, RSNNs have demonstrated effectiveness in learning and controlling complex behaviors \cite{lele2020learning,wang2023evolving}. They have also shown promise in path planning and navigation tasks \cite{rueckert2016recurrent,vieth2025stabilizing}, where their recurrent structure facilitates efficient spatial reasoning and decision-making.
Beyond these domains, RSNNs have been successfully applied to spatiotemporal tasks such as time-series forecasting \cite{lv2024efficient} and signal processing \cite{xu2024rsnn}, as detailed in Table~\ref{tab:rnn-app}. However, training RSNNs remains a principal bottleneck \cite{zhang2019spike}, as their temporal dynamics and recurrent structures require specialized computational techniques \cite{laddach2024adjusted} and hardware platforms \cite{yang202471,kim2025frame}. These challenges may hinder their practical deployment in resource-limited edge environments, underscoring the need for further research on efficient training strategies and scalable support for neuromorphic platforms.

\begin{table}[!t]
\centering
\caption{The summary of EdgeSNN applications with RNN topology.}
\label{tab:rnn-app}
\resizebox{\columnwidth}{!}{%
\begin{tabular}{lllr}
\toprule
\rowcolor[HTML]{f3fbf2}  
\textbf{Model} & \textbf{Year} & \textbf{Venue} & \textbf{Proposed Tasks} \\
\midrule \midrule
Improved SNN \cite{ponghiran2022spiking} & 2022 & AAAI & Speech Recognition \\
RSNN-SCBAM \cite{xu2023enhancing} & 2023 & NeurIPs & Event-based Image Classification \\ 
EAS-SNN \cite{wang2024eas} & 2024 & ECCV & Object Detection \\
EC-RSNN \cite{wang2023evolving} & 2023 & NeurIPs & Robotic Locomotion \\
SRSNN \cite{cherdo2023time} & 2023 & IJCNN & Time Series Prediction \& Anomaly Detection \\
Spike-RNN/GRU \cite{lv2024efficient} & 2024 & ICML & Time Series Prediction \\
CONLSTM \cite{xu2024rsnn} & 2024 & MM & Event-based Image Classification \\ 
SAD \cite{zhu2025autonomous} & 2024 & NeurIPs & Autonomous Driving \\ 
GRSN \cite{qin2025grsn} & 2025 & AAAI & Robotic Control \\
\bottomrule
\end{tabular}%
}
\end{table}

The remarkable success of transformer architectures in both NLP and CV tasks \cite{khan2022transformers} has inspired growing interest in adapting their capabilities to the SNN paradigm, aiming to achieve advanced learning performance under reduced energy budgets \cite{yao2023spike,shi2024spikingresformer}. Current research in this domain primarily follows two directions.
First, considerable effort has been devoted to reconciling the interior self-attention mechanism—a cornerstone of transformers—with the discrete, event-driven nature of SNNs. This is particularly challenging due to the incompatibility of softmax operations (which rely on exponentiation and division) with spike-based computation \cite{zhou2023spikformer}. To overcome these limitations, researchers have proposed several innovative alternatives, including spike-wise attention mechanisms that enhance feature representation \cite{guo2025spiking,lee2025spiking,zhang2025staa}, spike-friendly computational paradigm \cite{wang2023masked,xiao2025rethinking}, quantization-aware designs \cite{qiu2025quantized}, and memory-efficient architectures tailored for spike-driven systems \cite{zhang2024qkformer,liu2024lmuformer,deng2024spiking,yao2024spikedriven}. These techniques aim to bridge the gap between the continuous, dense operations of conventional transformers and the sparse, binary signaling of SNNs.
Secondly, the strong representational power of transformer-based models has been harnessed to improve SNN performance on more complex tasks. This has led to the development of transformer-inspired SNNs for a variety of applications, including multivariate time-series forecasting \cite{lv2024efficient}, language modeling \cite{xing2024spikelm,xing2024spikellm,zhu2024spikegpt}, signal processing \cite{guo2024spgesture,zhang2024spike}, reinforcement learning (RL) \cite{huang2025decision}, and advanced CV tasks \cite{lei2024spike2former,wang2025spiking,deng2024spiking}, as illustrated in Table~\ref{tab:tf-app}. Collectively, these advances underscore the potential of integrating transformer architectures with neuromorphic principles to develop more capable and energy-efficient edge intelligence. 

\begin{table}[!t]
\centering
\caption{The summary of Transformer-based EdgeSNN applications.}
\label{tab:tf-app}
\resizebox{\columnwidth}{!}{%
\begin{tabular}{lllr}
\toprule
\rowcolor[HTML]{f3fbf2}  
\textbf{Model} & \textbf{Year} & \textbf{Venue} & \textbf{Proposed Tasks} \\
\midrule \midrule
STNet \cite{zhang2022spiking} & 2022 & CVPR & Object Tracking \\
DyTr-SNN \cite{wang2023complex} & 2023 & AAAI & Speech Recognition \\ 
Spike Transformer \cite{zhang2022spike} & 2022 & ECCV & Depth Estimation  \\
MST \cite{wang2023masked} & 2023 & ICCV & Static \& Event-based Image Classification \\  
Spikformer \cite{zhou2023spikformer} & 2023 & ICLR & Static \& Event-based Image Classification \\ 
STS-Transformer \cite{wang2023spatial} & 2023 & IJCAI & Event-based Image \& Speech Classification \\  
Spike-Driven Transformer \cite{yao2023spike} & 2023 & NeurIPs & Static \& Event-based Image Classification \\ 
SpikingBERT \cite{bal2024spikingbert} & 2024 & AAAI & Natural Language Processing \\   
SpikingResformer \cite{shi2024spikingresformer} & 2024 & CVPR & Static \& Event-based Image Classification \\
SWformer \cite{fang2024spiking} & 2024 & ECCV & Static \& Event-based Image Classification \\ 
LMUFormer \cite{liu2024lmuformer} & 2024 & ICLR & Speech Recognition  \\ 
Meta-SpikeFormer \cite{yao2024spikedriven} & 2024 & ICLR &   Image Classification \& Segmentation \\  
OST \cite{song2024one} & 2024 & IJCAI & Static \& Event-based Image Classification \\  
TIM \cite{shen2024tim} & 2024 & IJCAI & Event-based Image Classification \\  
SpikeLM \cite{xing2024spikelm} & 2024 & ICML & Natural Language Processing \\
iSpikformer \cite{lv2024efficient} & 2024 & ICML & Time Series Prediction \\  
PSSD-Transformer \cite{wang2024pssd} & 2024 & MM & Image Segmentation \\ 
QKFormer \cite{zhang2024qkformer} & 2024 & NeurIPs & Static \& Event-based Image Classification \\ 
SpGesture \cite{guo2024spgesture} & 2024 & NeurIPs & Signal Processing \\
SSL \cite{zhang2024spike} & 2024 & NeurIPs & Signal Processing \\ 
STMixer \cite{deng2024spiking} & 2024 & NeurIPs & Static Image Classification \\ 
Spike2Former \cite{lei2024spike2former} & 2025 & AAAI & Image Segmentation \\  
SMA-AZO \cite{shan2025advancing} & 2025 & AAAI & Static \& Event-based Image Classification \\
SPT \cite{wu2025spiking} & 2025 & AAAI & 3D Reconstruction \\
FSTA-SNN \cite{yu2025fsta} & 2025 & AAAI & Static \& Event-based Image Classification \\ 
QSD-Transformer \cite{qiu2025quantized} & 2025 & ICLR & Static \& Event-based Image Classification \\ 
SpikeGPT \cite{zhu2024spikegpt} & 2025 & ICLR & Natural Language Processing \\ 
SNN-ViT \cite{wang2025spiking} & 2025 & ICLR & Image Classification \& Object Detection \\  
SpikeLLM \cite{xing2024spikellm} & 2025 & ICLR & Natural Language Processing \\
\bottomrule
\end{tabular}%
}
\end{table}

Integrating SNNs with other architectural paradigms, such as MLPs, has emerged as a promising approach for processing temporal data. Owing to their inherent ability to capture temporal dynamics efficiently \cite{kamata2022fully,li2022brain}, SNNs are well-suited for real-time applications, including motion analysis \cite{zhao2023recognizing}, dynamic scene reconstruction \cite{zhu2022event, liao2024spiking, zhu2024spikenerf,ren2024spikepoint}, and event-based pattern recognition \cite{skatchkovsky2021learning,xia2023unsupervised}. Moreover, the event-driven and low-latency nature of SNNs makes them particularly advantageous for continuous decision-making tasks, such as those involved in robotic control. Recent efforts have explored the integration of SNNs into RL frameworks \cite{qin2023low,wang2023event}, leveraging their computational efficiency to address the high resource demands traditionally associated with RL algorithms. However, many of these approaches adopt hybrid ANN-SNN architectures, which may face compatibility challenges when deployed on edge platforms equipped with conventional neuromorphic chips \cite{shen2016darwin,davies2018loihi}. Addressing these limitations will be crucial for realizing fully spiking implementations that can support real-time, on-device learning and decision-making.

In summary, SNNs with diverse architectural topologies have shown strong potential in tackling a wide range of complex computational tasks while maintaining intrinsic energy efficiency, making them particularly well-suited for edge computing and real-time processing scenarios.

\subsection{Learning Algorithm}

Enabling local training for SNN models on edge devices is crucial in many application scenarios. By eliminating data transmission, local training mitigates privacy risks and supports real-time adaptation to dynamic environments. It also enhances efficiency and sustainability in bandwidth-constrained or communication-cost-sensitive environments, such as those used in remote monitoring and smart device applications. These benefits underscore the importance of developing efficient, privacy-preserving, and cost-effective local training strategies for edge-deployed SNNs.

\begin{figure}[!t]
\centering
\includegraphics[width=\columnwidth]{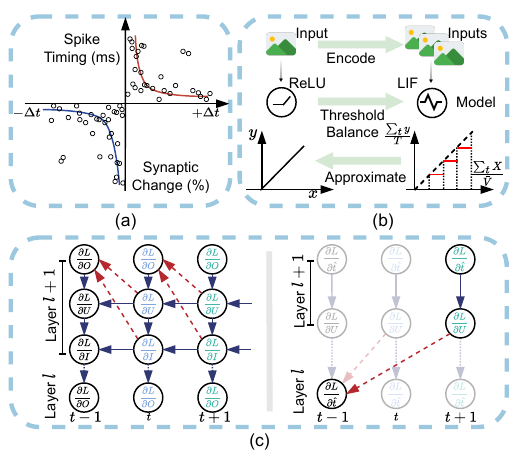}
\caption{Schematics of SNN learning algorithms. (a) Classic unsupervised STDP for synaptic modifications. (b) The conversion pipeline from pre-trained ANNs to SNNs. (c) Descriptions of rate coding-based (left) and temporal coding-based (right) direct training methods.} 
\label{fig:learnings}
\end{figure}

The training of SNNs is fundamentally a data-driven process that relies on learning algorithms to optimize network parameters for specific tasks. These algorithms play a pivotal role in determining model performance. As illustrated in Fig.~\ref{fig:learnings}, most currently effective SNN training methods are implemented on GPUs, which can be broadly categorized into two paradigms: \textbf{Unsupervised Learning}, inspired by biological synaptic plasticity, and \textbf{Supervised Learning}, which leverages techniques from deep learning. 

\subsubsection{Unsupervised Learning}
Synaptic plasticity forms the biological basis of neural learning, relying on the relative timing of spikes between connected neurons, as exemplified by Hebbian learning \cite{hebb2005organization}. Among various forms, spike-timing-dependent plasticity (STDP) is an asymmetric Hebbian rule that adjusts synaptic weights based on the precise temporal difference between presynaptic and postsynaptic spikes \cite{masquelier2008spike,fremaux2016neuromodulated}. As one of the most widely adopted bio-inspired unsupervised learning mechanisms in SNNs, STDP updates the synaptic weight $\Delta w_{ij}$ by the following rules:
\begin{align}
    \Delta w_{ij} = 
    \begin{cases}
        A^-\mathrm{exp}(\frac{t_i - t_j}{\tau^-}), \text{if } t_j - t_i \leq 0 \\
        A^+\mathrm{exp}(\frac{t_j - t_i}{\tau^+}), \text{if } t_j - t_i > 0
    \end{cases}
\end{align}
where $t_i$, $t_j$ denote the presynaptic and postsynaptic spiking timings, $\tau^+$, $\tau^-$ denote the constants affecting the scale of the time window, and $A^+$, $A^-$ correspond to long-term potentiation (LTP) and long-term depression (LTD), respectively.
Additionally, several STDP variants, such as mirrored STDP (mSTDP) \cite{burbank2015mirrored}, probabilistic STDP \cite{tavanaei2016acquisition}, and reward-modulated STDP (r-STDP) \cite{izhikevich2007solving,legenstein2008learning}, have been introduced to improve simulation speed and stabilize learning performance. 
While biologically plausible, these learning methods often confine SNNs to shallow architectures with limited representational capacity \cite{renner2024backpropagation,wu2022brain,safa2023fusing}. 
Although current STDP-based methods suffer from high computational costs and limited scalability, particularly on general-purpose hardware such as GPUs \cite{lv2025dendritic, journe2023hebbian}, their label-free nature remains attractive for EdgeSNN applications. As neuromorphic hardware continues to evolve, we believe these methods hold promise for enabling efficient, on-device adaptation in dynamic and resource-constrained IoT settings.

\subsubsection{Supervised Learning}
There are currently two main supervised learning paradigms for training SNNs: (1) \textbf{Direct Training} and (2) \textbf{ANN2SNN}.

The remarkable success of gradient descent (GD)-based algorithms in ANNs has inspired efforts to explore their applicability to the end-to-end training of SNNs in an efficient manner. Based on the neural coding schemes introduced in Section~\ref{sec:coding}, existing direct training methods for SNNs can be broadly classified into two categories: \textbf{Rate Coding} and \textbf{Temporal Coding}.
Rate coding-based methods address the non-differentiability of the spiking function by introducing surrogate gradients \textit{w.r.t.} spike activations. In contrast, temporal coding-based methods focus on spike timing and compute gradients \textit{w.r.t} the precise spike times \cite{kim2020unifying}.

In rate coding-based methods, Equation~(\ref{eq:update-mem}) can be explicitly extended to describe the membrane potential update of the $j$-th neuron as follows:
\begin{equation}
    \begin{aligned}
        & U_j[t]  = \tau \cdot H_j[t-1] + \sum_i w_{i,j} O_i[t] \\
        & \text{s.t. } O_i[t]  = \Theta(U_i[t]) = 
            \begin{cases}
                1, \text{if } U_i[t] \geq \bar{V}_i \\
                0, \text{o.w.} 
        \end{cases}
    \end{aligned}
\end{equation}
A surrogate function, or smooth activation function, is then employed as $\sigma(U[t]) = \frac{\partial O[t]}{\partial U[t]}$ to provide a continuous relaxation of the non-differentiable spiking behavior induced by the Heaviside step function $\Theta(\cdot)$. This relaxation enables the use of standard backpropagation through time (BPTT) to train SNNs from scratch \cite{guo2024take,yu2024advancing,lian2023learnable,wang2023adaptive,li2021differentiable,perez2021sparse}.
Specifically, this direct training method differs in the functional forms of surrogate gradients it employs, reflecting diverse strategies for approximating spiking nonlinearity. 
For example, SML \cite{deng2023surrogate} constructs a shortcut path to back-propagate a more accurate gradient to a particular SNN part. DeepTAGE \cite{liu2025deeptage} dynamically adjusts surrogate gradients following the membrane potential distribution across different time steps, enhancing their respective gradients in a temporally aligned manner that promotes balanced training.
STC \cite{wang2024autaptic} integrates two learnable adaptive pathways, enhancing the spiking neurons’ temporal memory and spatial coordination and mitigating the issue of gradient vanishing.
In \cite{wu2019direct,wu2018spatio}, a spatiotemporal backpropagation considering both the spatial and temporal credit assignment in roll-out SNNs based on both the iterative LIF model and several approximated derivatives of spike activities.

Regarding temporal coding-based methods, a spike response kernel is often employed to characterize the influence of a spike emitted by one neuron on its postsynaptic counterparts. This formulation allows for the simulation of SNNs without explicit membrane potential integration, treating spike timings as the primary state variables rather than binary firing indicators. Under this perspective, Equation~(\ref{eq:update-mem}) can be reformulated as:
\begin{align}
    U_j[t] = \sum_i\sum_{\hat{t}_i \in T_{i,j,t}} w_{i,j}\epsilon[t-\hat{t}_i]
\end{align}
where $\hat{t}_i \in T_{i,j,t}=\{\hat{t}_i | \hat{t}^{last}_j < \hat{t}_i < t, O_i[\hat{t}_i]=1 \}$ denotes the spiking times of the $i$-th afferent neuron, and $t^{last}_i$ is the last spiking time of $j$-th neuron before $t$. 
The derivative of the spiking time \textit{w.r.t.} the membrane potential, $\sigma(U_j[t])=\frac{\partial \hat{t}_j}{\partial U_j[\hat{t}_j]}$, is incorporated into the backpropagation chain, named temporal learning-based backpropagation (TLBP). 
TLBP restricts gradient propagation to discrete spike times, unlike the rate coding-based methods, which propagate gradients throughout the entire time window. This temporally sparse computation not only reduces unnecessary updates but also highlights the potential for improved energy efficiency in training SNNs.
Among existing studies, SpikeProp \cite{bohte2002error} is a pioneering work in this category, where the spike times of hidden units are linearized to enable analytical gradient approximation in hidden layers. 
Rate-based loss functions are mapped to their time-based counterparts, thereby enabling their use in time-based training schemes and revealing the implicit relationship between rate and temporal coding in SNNs \cite{zhu2023exploring}.
In \cite{zhu2022training}, a backward kernel function is proposed to address the reverse gradient problem in the event-driven learning process.

In practice, rate coding-based training methods are more suitable for deployment on general-purpose edge devices, as their reliance on BPTT can be effectively accelerated by GPUs. In contrast, although temporal coding-based learning rules offer the potential for highly energy-efficient SNN training, they currently require specialized neuromorphic hardware to achieve practical training speed \cite{zhang2024anp}. When executed on GPUs, such methods may even incur higher energy costs \cite{kim2020unifying}. This situation suggests that rate coding-based methods remain more viable for training EdgeSNNs in the near term. However, we posit that with the continued advancement of neuromorphic chips, temporal coding-based learning rules may emerge as a dominant paradigm for training SNNs in the future.

\begin{table}[!t]
\centering
\caption{Summary of TOP-10 SNN learning algorithm performance on common image classification benchmarks.}
\label{tab:bench-algo}
\resizebox{\columnwidth}{!}{%
\begin{threeparttable}
\begin{tabular}{lllrrr}
\toprule
\textbf{Ref.} & \textbf{Year} & \textbf{Venue} & \textbf{Learning} & \textbf{Network}$^\dagger$ & \textbf{Accuracy$^\ddagger$ (\%)} \\ 
\midrule \midrule
\rowcolor[HTML]{f3fbf2}  
\multicolumn{6}{c}{\textbf{CIFAR10}}  \\
\midrule
\cite{you2024spikezip} & 2024 & ICML & ANN2SNN & ViT & 98.70 \\
\cite{shi2024spikingresformer}  & 2024 & CVPR & Direct Train & ViT & 97.40 \\ 
\cite{hwang2024spikedattention} & 2024 & NeurIPs & ANN2SNN & ViT & 97.30 \\ 
\cite{wang2023masked} & 2023 & ICCV & ANN2SNN & ViT & 97.27 \\ 
\cite{zhang2025staa} & 2025 & CVPR & Direct Train & ResNet & 97.14 \\
\cite{deng2023surrogate} & 2023 & ICML & Direct Train & ResNet & 96.82 \\ 
\cite{huang2024clif} & 2024 & ICML & Direct Train & ResNet & 96.69 \\ 
\cite{li2022efficient} & 2022 & IJCAI & ANN2SNN & ResNet & 96.59 \\ 
\cite{guo2022reducing} & 2022 & ECCV & Direct Train & ResNet & 96.49 \\ 
\cite{qiu2024gated} & 2024 & AAAI & Direct Train & ResNet & 96.46  \\
\midrule
\rowcolor[HTML]{f3fbf2}  
\multicolumn{6}{c}{\textbf{CIFAR100}} \\
\midrule
\cite{you2024spikezip} & 2024 & ICML & ANN2SNN & ViT & 89.70  \\
\cite{wang2023masked} & 2023 & ICCV & ANN2SNN & ViT & 86.91 \\
\cite{hwang2024spikedattention} & 2024 & NeurIPs & ANN2SNN & ViT & 86.30 \\
\cite{shi2024spikingresformer} & 2024 & CVPR & Direct Train & ViT & 85.98 \\
\cite{guo2024enof} & 2024 & NeurIPs & ANN2SNN & ResNet & 82.43 \\
\cite{zhang2025staa} & 2025 & CVPR & Direct Train & ResNet & 82.05 \\
\cite{deng2024spiking} & 2024 & NeurIPs & Direct Train & ViT & 81.87 \\
\cite{deng2023surrogate} & 2023 & ICML & Direct Train & ResNet & 81.70 \\
\cite{hao2023progressive} & 2023 & ICLR & Direct Train & ResNet & 81.65 \\
\cite{xu2024bkdsnn} & 2024 & ECCV & ANN2SNN & ViT & 81.26 \\
\midrule
\rowcolor[HTML]{f3fbf2}  
\multicolumn{6}{c}{\textbf{ImageNet}} \\
\midrule
\cite{you2024spikezip} & 2024 & ICML & ANN2SNN & ViT & 83.82 \\
\cite{lee2025spiking} & 2025 & CVPR & Direct Train & ViT &  80.67 \\
\cite{hwang2024spikedattention} & 2024 & NeurIPs & ANN2SNN & ViT & 80.00  \\
\cite{shi2024spikingresformer} & 2024 & CVPR & Direct Train & ViT &  79.40 \\
\cite{li2021free} & 2021 & ICML & ANN2SNN & ResNet & 79.21 \\
\cite{guo2025spiking} & 2025 & CVPR & Direct Train & ViT & 78.66 \\
\cite{wang2023masked} & 2023 & ICCV & ANN2SNN & ViT &  78.51 \\
\cite{wang2025adaptive} & 2025 & AAAI & ANN2SNN & ViT & 77.09 \\
\cite{shan2025advancing} & 2025 & AAAI & Direct Train & ResNet & 77.05 \\
\cite{deng2024spiking} & 2024 & NeurIPs & Direct Train & ViT & 76.68 \\
\midrule
\rowcolor[HTML]{f3fbf2}  
\multicolumn{6}{c}{\textbf{CIFAR10-DVS}} \\
\midrule
\cite{you2024spikezip} & 2024 & ICML & ANN2SNN & ViT & 90.50 \\
\cite{shen2024rethinking} & 2024 & NeurIPs & Direct Train & VGG &  87.80 \\
\cite{huang2024clif} & 2024 & ICML & Direct Train & VGG &  86.10 \\
\cite{fang2023parallel} & 2023 & NeurIPs & Direct Train & VGG & 85.90 \\
\cite{deng2023surrogate} & 2023 & ICML & Direct Train & ResNet &  85.23 \\
\cite{shi2024spikingresformer} & 2024 & CVPR & Direct Train & ViT &  84.80 \\
\cite{wang2023adaptive} & 2023 & ICML & Direct Train & VGG &  84.50 \\
\cite{zhang2024qkformer} & 2024 & NeurIPs & Direct Train & ViT &  84.00 \\
\cite{shan2025advancing} & 2025 & AAAI & Direct Train & VGG & 84.00 \\
\cite{fang2024spiking} & 2024 & ECCV & Direct Train & ViT &  83.90 \\
\midrule
\rowcolor[HTML]{f3fbf2}  
\multicolumn{6}{c}{\textbf{DVS-Gesture}} \\
\midrule
\cite{huang2024clif} & 2024 & ICML & Direct Train & ViT  &  99.31 \\
\cite{zhou2024spiking} & 2024 & NeurIPs & Direct Train & ViT &  99.30 \\
\cite{yao2023spike} & 2023 & NeurIPs & Direct Train & ViT &  99.30 \\
\cite{song2024one} & 2024 & IJCAI & Direct Train & ViT &  99.00 \\
\cite{kachole2024asynchronous} & 2024 & ECCV & Direct Train & MLP &  98.74 \\
\cite{wang2023spatial} & 2023 & IJCAI & Direct Train & ViT &  98.72 \\
\cite{meng2023towards} & 2023 & ICCV & Direct Train & VGG &  98.62 \\
\cite{zhang2025staa} & 2025 & CVPR & Direct Train & VGG & 98.61 \\
\cite{zhang2024qkformer} & 2024 & NeurIPs & Direct Train & ViT &  98.60 \\
\cite{shan2025advancing} & 2025 & AAAI & Direct Train & VGG & 98.60 \\
\bottomrule
\end{tabular}%
\begin{tablenotes}
\item[$\dagger$] The network topology is all spike-wise.
\item[$\ddagger$] Results are directly taken from the original studies.
\end{tablenotes}
\end{threeparttable}
}
\end{table}

ANN2SNN is another effective alternative to training high-performance SNNs, particularly for applications oriented towards ANNs. The core idea is to approximate the continuous activations of ANNs with ReLU nonlinearity using the average firing rates of spiking neurons \cite{cao2015spiking,ijcai2021p321}.
With appropriate weight adjustments, a trained ANN can be converted into a spiking counterpart that preserves a similar input–output mapping. This approach leverages backpropagation in the ANN, thereby circumventing the non-differentiability issue introduced by the Heaviside function in Equation~(\ref{eq:heaviside}) when applying gradient-based learning directly to SNNs. The resulting SNNs incur minimal accuracy loss compared to their ANN counterparts, which can be effectively extended to a wide range of architectures, including VGG, ResNet \cite{hao2023bridging,bu2022optimal1}, and large-scale Vision Transformers (ViTs) \cite{you2024spikezip,huang2024towards,hwang2024spikedattention}.
However, the approximation process of ANN2SNN overlooks the intrinsic spatiotemporal dynamics of SNNs, thereby limiting their representational capacity in event-driven scenarios \cite{deng2020rethinking}. 
Moreover, ANN2SNN typically requires long simulation periods, leading to substantial computational and memory overhead \cite{li2024brain}, which makes it unsuitable for direct deployment on resource-constrained edge devices.
However, we believe that with the support of resource-rich cloud servers and cloud–edge collaborative training framework \cite{yao2022edge}, ANN2SNN may become a viable solution for enabling SNN training at the edge.

Compared to direct training methods, ANN2SNN approaches often involve additional computational overhead and complex calibration procedures, which limit their practicality for intuitive deployment on resource-constrained edge devices. 
Nevertheless, the computational capacity of cloud servers might enable a collaborative ANN2SNN framework, in which heavy processing is handled in the cloud, and only affordable, time-critical operations are performed locally, offering a practical pathway for deploying ANN2SNN in edge scenarios.

\subsubsection{Hybrid Learning}
While spike-based gradient methods achieve high accuracy, they suffer from substantial memory overhead due to the need for end-to-end backpropagation. In contrast, STDP-based learning rules are memory-efficient but often fall short in performance. To balance accuracy and efficiency, hybrid learning strategies that integrate both paradigms have been actively explored. EIHL \cite{jiang2024adaptive} dynamically adjusts network connectivity through an excitation-inhibition mechanism, enabling adaptive switching between STDP-based local learning and gradient-based global learning. Similarly, EICIL \cite{shao2023eicil} proposes an iterative training scheme that combines STDP and surrogate gradient methods to enhance the expressiveness of spiking neurons and overall SNN performance.
SpikeSlicer \cite{cao2024spiking} proposes an ANN-SNN cooperative paradigm that adopts a feedback updating strategy, refining SNNs using feedback from downstream ANNs.
These examples highlight the potential of hybrid learning strategies to meet edge computing constraints by combining memory efficiency with practical training, making them a compelling avenue for future research.

Table~\ref{tab:bench-algo} also summarizes the comparative studies about all these learning algorithms.

\subsection{Hardware Support}
\label{sec:hardware}

Hardware platforms provide the computational foundation for EdgeSNNs.
Efficient hardware can significantly accelerate the evaluation of SNN-based models, facilitating the exploration of novel architectures and their real-world applications.
However, as an emerging technology, no commercially available neuromorphic hardware can support all EdgeSNN applications.
Consequently, most neuromorphic research focuses on algorithmic advancements using conventional hardware, such as GPUs. While not optimal for performance, these platforms can serve as effective simulation environments for validating system functionality and providing valuable insights to guide future neuromorphic hardware optimizations.
Below, we introduce two currently developed BIC-oriented hardware types: neuromorphic sensors and chips.

\begin{figure}[!t]
\centering
\includegraphics[width=\columnwidth]{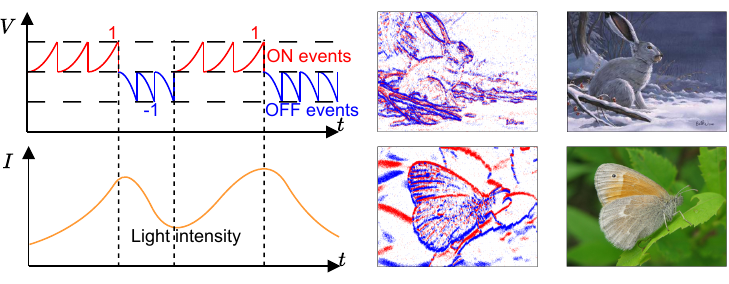}
\caption{Comparison of neuromorphic and conventional visual sensors (adapted from \cite{kim2021n}). The frames transformed from raw event streams are much sparser than conventional RGB images.} 
\label{fig:cmp-imgs}
\end{figure}

\subsubsection{Neuromorphic Sensor}

Neuromorphic sensors are advanced devices designed to emulate the sensory processing capabilities of biological organs, including the retina, cochlea, and skin, by leveraging principles from neuromorphic engineering. These sensors process information in a manner analogous to the human nervous system, enabling the rapid and efficient handling of sensory data \cite{liu2010neuromorphic}.

Since most existing studies on SNN-based applications primarily focus on the \textbf{vision} domain, the development of neuromorphic sensors has mainly centered on neuromorphic cameras, which use spatiotemporal one-bit points to encode light intensity \cite{posch2014retinomorphic}. 
One notable example is the event camera \cite{gallego2020event}, namely the silicon retina. 
Representative event cameras include DVS128 \cite{lichesteiner2008128}, DAVIS \cite{li2015design}, Celex \cite{guo2017live}, ATIS \cite{simon2016event}, SciDVS \cite{graca2024scidvs}, etc.
These innovative imaging devices, inspired by the human retina, are designed to detect changes in brightness at the pixel level asynchronously.
Unlike traditional frame-based cameras that capture entire scenes at fixed intervals, event cameras respond only to variations in illumination, resulting in lower data rates, reduced latency, higher dynamic range, and lower power consumption, as illustrated in Fig.~\ref{fig:cmp-imgs}. 
These attributes make DVS suitable for applications requiring rapid response times and efficient data processing \cite{trivedi2025intelligent}.
So far, event cameras have been the mainstream neuromorphic vision sensors in academic research and industrial applications in robotics and autonomous systems, including pattern recognition \cite{yang2023event,liu2024seeing,jiang2024evhandpose,safa2023fusing}, object detection \cite{gehrig2023recurrent,wu2024leod,cao2024chasing}, segmentation 
\cite{zhu2024continuous, li2024event, jing2024hpl, kong2024openess}, depth estimation \cite{shiba2024secrets,liu2024event,shi2023even}, and optical flow estimation \cite{wu2024lightweight, shiba2024secrets, zhao2024optical}, mainly due to the well-established ecosystem fostered by open-source datasets, code, and software tools \cite{zheng2023deep}. 

\begin{figure*}[!t]
\centering
\includegraphics[width=\textwidth]{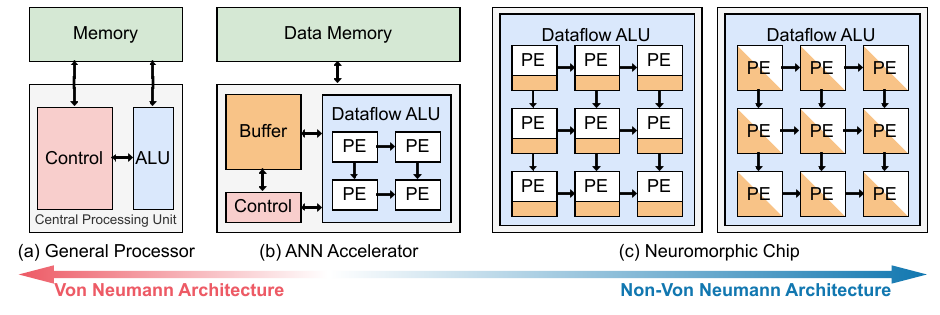}
\caption{Architecture comparison of (a) general processors, (b) ANN accelerators, and (c) neuromorphic chips featuring decentralized many-core architectures with either (left) near-memory computing or (right) in-memory computing (Adapted from \cite{shin2019heterogeneous}).} 
\label{fig:chip-archs}
\end{figure*}

Other neuromorphic sensors are also designed to emulate the sensory processing capabilities of the human nervous system, extending to a broader range of sensing modalities.

In the \textbf{auditory} domain, sensors like Xylo Audio \cite{bos2023sub} mimic the human cochlea's ability to decompose complex sounds into their frequency components, enhancing sound perception and processing in technological applications. 
These neuromorphic auditory sensors offer real-time, low-power, and efficient auditory data processing, making them suitable for speech recognition \cite{nilsson2023comparison,jimenez2016binaural,alsakkal2025spiketrum}, localization \cite{haghighatshoar2025low,zhang2024spike,zheng2023neuroradar}, and environmental sound analysis \cite{lenk2023neuromorphic}. 

Additionally, neuromorphic \textbf{tactile} sensors \cite{macdonald2022neuromorphic} mimic the human sense of touch, thereby enhancing the dexterity and object manipulation capabilities of robots and prosthetic devices. For instance, the NeuroTac sensor \cite{ward2020neurotac} integrates bio-mimetic hardware with an event-based camera to convert tactile information into spike trains, enabling texture-recognition tasks \cite{faris2024novel}. Similarly, multilayered tactile sensors \cite{sankar2025natural,liu2023progress} inspired by human skin have been developed, allowing for compliant grasping and the differentiation of surface textures in everyday objects.

For Neuromorphic \textbf{olfactory} sensors \cite{birkoben2020spiking}, they emulate the biological olfactory system to detect and process odor information efficiently, enabling applications in environmental monitoring—such as detecting hazardous gases and pollutants—and in the food and beverage industry for quality control through aroma profiling \cite{tayarani2021event,jang2024autonomous}. For instance, an SNN-based classifier \cite{vanarse2022application} has demonstrated high accuracy in distinguishing different malts based on their scent profiles.

Due to their event-driven nature, neuromorphic sensors hold great promise for edge computing applications. Their efficiency and versatility are particularly beneficial for edge devices that require continuous data processing under stringent power constraints. When combined with SNNs, which are inherently adept at handling temporal information, this synergy facilitates the development of adaptive, low-power systems suited for real-time, bio-inspired processing tasks \cite{rathi2023exploring,niedermeier2025integrated}. Such integration could significantly enhance the intelligence and responsiveness of edge computing infrastructure. However, realizing this potential poses notable challenges, including the difficulty of replicating biological neural dynamics in hardware, the shift from synchronous to asynchronous programming models, and the lack of standardized evaluation frameworks \cite{schuman2022opportunities}.

\subsubsection{Neuromorphic Chip}

Deep learning chips are designed to optimize the execution of ANNs, achieving better power-performance-area (PPA) efficiency than general-purpose processors. These chips typically follow von Neumann architectures with sophisticated processing units and hierarchical memory structures \cite{shin2019heterogeneous}. They leverage data reuse in convolution operations by employing data flow architectures that maximize reuse rates between processing elements (PEs). Their memory hierarchy resembles CPUs/GPUs, incorporating off-chip DRAM and multi-level on-chip buffers.

While deep learning accelerators prioritize performance, neuromorphic chips emphasize efficiency. They are designed explicitly for SNN emulation and adopt non-von Neumann, decentralized many-core architectures with tightly coupled computation and memory resources. 
Fig.~\ref{fig:chip-archs} provides several kinds of common architectures of neuromorphic chips.
Neuromorphic chips focus on massive parallelism, high memory locality, and strong scalability, extending from individual cores to entire chips and even multi-board systems \cite{basu2022spiking}. 

From a \textbf{functionality} perspective, a chip's capabilities are determined by its instruction set and the models it supports. Early neuromorphic chips support SNNs as the primary model, while several chips have expanded their support to include ANNs for execution efficiency. 
Additionally, learning ability is a crucial functionality that will be further discussed.

Early neuromorphic chips, inspired by the concept of neuromorphic engineering \cite{mead2012analog}, were primarily designed to support brain-inspired SNNs. Notable examples include Darwin \cite{ma2024darwin3}, Loihi \cite{davies2018loihi}, and DYNAPs \cite{moradi2017scalable}, which have been widely applied in domains such as speech and image processing. Due to the event-driven nature and sparse activity of SNNs, neuromorphic chips generally achieve significantly lower power consumption than conventional processors when executing SNN workloads. This efficiency is primarily attributed to compute gating and even clock gating techniques. Additionally, latency can be further reduced by processing only validated spike events while skipping zero events \cite{eshr2023train}, enhancing computational efficiency.
Although supporting SNNs to enable energy-efficient and powerful intelligent applications at the edge is promising, it cannot be denied that research on SNNs is still in its early stages of development. Researchers continue to explore ways to incorporate more brain-inspired features into SNN modeling to enhance its performance in real-world tasks, despite SNNs already demonstrating breakthroughs in various domains. However, this does not conceal the current dilemma: SNNs have yet to demonstrate superior results beyond execution efficiency compared to ANNs \cite{deng2020rethinking}. Several modern neuromorphic chips, such as the Tianji Series \cite{pei2019towards,deng2020tianjic}, are designed with a cross-paradigm approach to bridge this gap. By integrating the high accuracy of ANNs with the rich dynamics, high efficiency, and robustness of SNNs, these chips pave the way for advancing brain-inspired edge systems toward higher levels of intelligence.

Moreover, given that learning capability is a fundamental brain function, recent studies have aimed to equip neuromorphic hardware with dedicated on-chip learning mechanisms for SNNs, enabling models to mitigate performance degradation in dynamically evolving IoT environments.
However, most existing neuromorphic chips are primarily designed for the inference phase of SNNs, necessitating the learning process being conducted on GPUs in advance. This approach introduces additional complexity and increases power consumption \cite{deng2023surrogate, jiang2018design}, as GPUs are optimized for ANN-oriented workloads rather than neuromorphic learning tasks.
Unlike gradient descent, STDP updates synaptic weights locally, avoiding costly weight update routing and reducing chip area overhead. This advantage has motivated researchers to develop neuromorphic chips that support the STDP learning mechanism \cite{vohra2024circuit,vohra2023full,davies2018loihi}, enabling energy-efficient in-situ SNN training without compromising parallelism \cite{diehl2015unsupervised, nessler2013bayesian}.
Meanwhile, prototype chips \cite{liu202382,wei2024event,zhong2023efficient}, such as ANP-I \cite{zhang2024anp}, also attempt to support temporal-based learning with lower energy overhead for edge-AI applications.

In summary, neuromorphic chips achieve efficient, low-power computing by mimicking the neural network architecture of the human brain, making them well-suited for resource-constrained edge devices. 
Nevertheless, their development is complex and time-consuming, and they are often optimized for specific tasks, which limits their general capability. Moreover, compared to traditional computing architectures, the development tools, programming models, and software support for neuromorphic computing remain underdeveloped, posing challenges to its widespread adoption in edge computing applications.

\section{Practical Considerations for EdgeSNNs}
\label{sec:apps} 

\begin{figure*}[!t]
\centering
\includegraphics[width=0.85\textwidth]{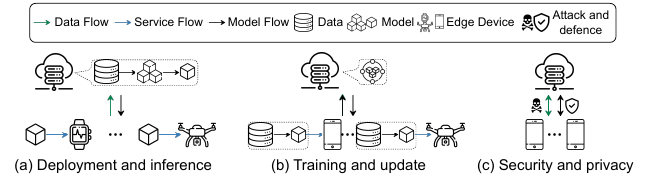}
\caption{Illustration of different practical considerations for EdgeSNN applications, including: (a) on-device deployment and inference, (b) training and update strategies, and (c) security and privacy risks along with countermeasures during system implementation.} 
\label{fig:apps}
\end{figure*} 

Fig.~\ref{fig:apps} presents a systematic overview of the practical considerations for EdgeSNN applications, which are further elaborated in the subsequent three sections.

\subsection{Deployment and Inference for EdgeSNNs}

This section introduces the deployment and inference strategies for EdgeSNNs, aiming to compress deep SNN models to fit the resource limitations of edge devices.
Recent developments have shifted toward deploying deep SNNs, trained initially in resource-abundant cloud environments, on memory-limited edge devices \cite{ijcai2024p596,yu2024fedlec}. This transition enables the provision of sophisticated, intelligent services at the edge, with potential energy efficiency gains due to the sparse and event-driven nature of SNNs. 
While some model compression methods are not explicitly proposed for EdgeSNNs in their original publications \cite{na2022autosnn,liu2024lite}, they inherently exhibit potential for deployment on resource-constrained devices. 
Accordingly, we review various general SNN model compression techniques, emphasizing their applicability and possible advantages for device-side deployment. 
These methods \cite{shi2024towards,shen2025improvinging, wei2024q} can be broadly categorized into: (1) \textbf{Lightweight Model Design} seeks to design more efficient network architectures from scratch with edge deployment constraints in mind; (2) \textbf{Post-hoc Model Simplification} refers to the use of lightweight techniques to simplify pre-trained SNN models, making them suitable for edge deployment. The latter category can be further subdivided into model compression and precision reduction, each targeting different aspects of model simplification.

\subsubsection{Lightweight Model Design}
This category encompasses techniques that aim to reduce model size and computational overhead by designing compact and efficient architectures from scratch, with minimal compromise in predictive performance. 
Neural Architecture Search (NAS) \cite{kim2022neural,na2022autosnn} is an automated approach for identifying architectures that offer an optimal trade-off between performance and efficiency.
Lite-SNN \cite{liu2024lite} proposes a joint optimization framework that simultaneously learns architecture parameters and optimizes time steps, aiming to achieve high-performance SNNs while minimizing memory usage and computational cost.
SpikeDHS \cite{che2022differentiable} identifies effective SNN architectures with low computational overhead by performing NAS at both the neuron cell level and the layer level.
ESL-SNN \cite{shen2023esl} dynamically learns synaptic connections in SNNs during training, maintaining structural sparsity to facilitate lightweight computation. The lottery ticket hypothesis has been applied to SNNs \cite{kim2022exploring}, aiming to discover sparse sub-networks—so-called "winning tickets"—that retain accuracy comparable to the original dense architectures.

Apart from automated NAS approaches, some studies have pursued human-invented lightweight SNNs by converting or adapting compact ANN architectures, including spiking versions of SqueezeNet and MobileNet \cite{cordone2022object,li2021free}.
Nevertheless, the performance of these shallow spiking variants tends to degrade in the presence of complex or high-dimensional data, limiting their applicability in real-world scenarios. To address this issue, researchers have introduced \textbf{knowledge distillation} \cite{xu2023constructing,yu2025temporal}, which transfers feature representations from a pre-trained ANN (teacher) to an SNN (student) to enhance the latter’s performance. 
BKDSNN \cite{xu2024bkdsnn} employs randomly blurred features from the SNN to facilitate knowledge distillation from the ANN, thereby reducing the performance gap between the two models. A multi-modal synergistic distillation scheme \cite{wang2024apprenticeship} allows a lightweight uni-modal student SNN to learn rich representations from an event-frame teacher network. HSD \cite{zhong2024towards} proposes a step-wise knowledge distillation strategy to reduce the adverse impact of anomalous outputs at individual time steps on the temporal average output of SNNs. 
Integrating knowledge distillation and connection pruning enables dynamic optimization of synaptic connectivity in SNNs, facilitating task-specific performance improvements \cite{xu2024reversing}. Inspired by ANN2SNN, a distillation framework based on block-wise replacement \cite{yang2025efficient} is proposed to align SNN feature spaces with those of ANNs using rate-based representations, thereby achieving comparable or even superior performance.

\subsubsection{Post-Hoc Model Simplification}
\textbf{Pruning} is a widely used compression approach that reduces model complexity by eliminating redundant computations or connections. It facilitates the deployment of prepared deep SNNs on resource-limited edge devices. In line with conventional pruning approaches, structured pruning has been applied to SNNs, aiming to eliminate unnecessary channels by evaluating their spiking activity \cite{ijcai2024p596,li2024towards}. This approach is hardware-friendly because it aligns with parallel processing and memory access patterns. 
In contrast, unstructured pruning \cite{chen2022state,xu2024reversing,ijcai2021p0236,zhuge2024towards} focuses on eliminating individual synapses, producing sparse SNNs that require fewer computational resources.
To leverage both advantages, a hybrid pruning strategy is proposed in \cite{shi2024towards}, eliminating unimportant neurons and synapses to maximize the energy efficiency of SNNs while maintaining acceptable accuracy. 

Unlike ANNs, SNNs introduce an additional pruning target: the temporal dimension. Accordingly, temporal pruning methods \cite{zhong2024towards,chowdhury2022towards,chakrabortysparse,ding2024shrinking} have been proposed to shorten the number of inference time steps, thereby reducing latency and energy consumption. 
DCT-SNN \cite{garg2021dct} employs the discrete cosine transform (DCT) to reduce the number of timesteps required for inference, achieving a 2× to 14× reduction in latency.
SEENN-II \cite{li2023seenn} introduces a reinforcement learning-based mechanism to determine optimal early-exit points during SNN inference, thereby reducing latency while maintaining accuracy.
Several studies \cite{li2023unleashing,li2023seenn,you2025vistream} also utilize a confidence-based criterion to evaluate input difficulty, thereby determining an adaptive number of time steps required for accurate inference on a per-sample basis.

To achieve a trade-off between model performance and memory cost, numerical precision reduction (a.k.a., \textbf{Quantization}) replaces 32-bit full-precision weights with multi-bit integer representations selected from a predefined set of quantized values. This alleviates the accuracy loss commonly associated with rigid mappings \cite{yang2019quantization}.
In contrast to quantization strategies in ANNs, which often involve both weight and activation quantization, SNN quantization techniques mainly target model-specific parameters, including synaptic weights and membrane potentials. Since inter-layer signals in SNNs are naturally represented as binary spikes, additional activation quantization is unnecessary, making them inherently memory-efficient.
QSD-Transformer \cite{qiu2025quantized} introduces quantized spike-driven self-attention to reduce the substantial computational resources from SNNs with large-scale transformer structures.
Meanwhile, quantifying synaptic weights and membrane potentials with different bit widths has been successfully applied to ResNet- and VGG-like SNNs in \cite{wei2025qp,wei2024q}, demonstrating a minimal impact on model accuracy.

\subsection{Training and Update for EdgeSNNs}

This section introduces training and update strategies for EdgeSNNs, with a focus on enabling deployed models to adapt to data generated during on-device operation continuously. Rather than treating edge learning as a one-time local training process, we emphasize continual and adaptive updates to maintain responsiveness to non-stationary and locally evolving data distributions. These strategies enhance privacy and security by storing sensitive data locally on the device, while also enabling the model to tailor its behavior to the local context.
However, limited computational resources and scarce labeled data on edge devices pose significant challenges to the effectiveness of local updates \cite{xu2021edge}. As a result, SNN training paradigms must be restructured to accommodate edge-specific constraints, motivating a shift toward collaborative and distributed learning frameworks that address data heterogeneity, resource limitations, and communication overhead. This shift, in turn, introduces new challenges, including reduced training and communication efficiency, heightened privacy risks, and increased uncertainty in model behavior due to the data's fragmentation and dynamic nature.

\subsubsection{Isolate Training and Update} 
The feasibility of training SNNs on a single device largely depends on how well the computational demands of the selected training algorithm align with the device's hardware constraints. Accordingly, this section discusses:
(1) strategies for efficiently accelerating SNN training on a single device, and
(2) mechanisms for preserving the learning plasticity of SNNs in response to dynamic local data distribution shifts at the edge.

For \textbf{training acceleration}, the study in \cite{zheng2023spiking} illustrates the practicality of on-device SNN training by verifying the effectiveness of STDP learning and its compatibility with integrated circuit implementations. Stochastic temporal flexibility \cite{du2025temporal,anumasa2024enhancing} is introduced to reduce the training cost of SNNs by adaptively adjusting temporal parameters during training. A token sparsification technique incorporating a timestep-wise anchor token and dual alignment strategies was proposed in \cite{zhuge2024towards} to enable efficient training of Spiking Transformers. MPIS-SNNs \cite{cao2024efficient} drive multiple fused parallel implicit streams to reach equilibrium simultaneously, thereby avoiding the need to store a large number of intermediate activations during backpropagation. SLTT-K \cite{meng2023towards} reduces memory consumption by ignoring unimportant routes in the computational graph during backpropagation, thereby decreasing the number of scalar multiplications and achieving memory usage that is independent of the limited number of time steps $K$ $(K<T)$.

To guarantee \textbf{learning plasticity}, HLOP \cite{xiao2024hebbian} offers valuable insights into leveraging neural circuit mechanisms and Hebbian learning to enhance the efficiency of incremental learning in SNNs, thereby paving the way for the development of low-cost, continual neuromorphic computing systems.
SA-SNN \cite{shen2024efficient} offers a feasible solution for augmenting continual learning in machine intelligence under limited computational resources by integrating neuromorphic hardware.
ALADE-SNN \cite{ni2025alade} introduces an efficient continual learning strategy for SNNs that addresses the issue of catastrophic forgetting in dynamic task scenarios, with a computational footprint suitable for edge device deployment.
DSD-SNN \cite{han2023enhancing} enables deployment-friendly continual learning by dynamically growing neurons for new tasks and pruning redundant ones, optimizing memory and computation for edge devices.

\subsubsection{Collaborative Training and Update}
The \textbf{master-slave} architecture is a widely adopted collaborative training paradigm, with federated learning (FL) \cite{mcmahan2017communication} being a representative example. In this framework, a central server orchestrates training by distributing tasks to multiple edge devices, which perform local updates and periodically communicate model parameters or gradients back to the server for aggregation.
In \cite{venkatesha2021federated}, this collaborative training framework is applied to SNNs for the first time by incorporating the BPTT algorithm. This integration demonstrates the feasibility of on-device deployment, as the learning process can be tailored to accommodate the limited computational and memory resources of edge devices \cite{hu2024high,yu2024advancing}. 
As a follow-up study, FedLEC \cite{yu2024fedlec} investigates the impact of label skewness in federated SNN learning frameworks and proposes a distribution calibration combined with a distillation strategy to mitigate its adverse effects.
FL-SNN \cite{skatchkovsky2020federated} generalizes the STDP rules to FL settings, enabling a flexible trade-off between communication overhead and accuracy in training on-device SNNs.
AdaFedAsy-SNN \cite{wang2023efficient} introduces an asynchronous federated SNN learning framework that employs a weight aggregation strategy based on average spike rates, effectively addressing the challenges posed by non-iid data distributions across devices.

Although FL has been widely adopted in collaborative training scenarios, its limitations in handling device heterogeneity and communication latency have become increasingly apparent in real-world deployments. To overcome these challenges, cloud-edge collaborative architectures \cite{yao2022edge} have been proposed, emphasizing the synergy between edge nodes and cloud servers and providing a more hierarchical and adaptive training paradigm. 
ECC-SNN \cite{yu2025eccsnn} pioneers the integration of on-device adaptation and cloud-guided supervision for SNNs, where edge devices update locally with support from a cloud-based pre-trained model. This design enhances responsiveness in dynamic edge environments. 

Another form of collaborative learning architecture is the \textbf{peer-to-peer} paradigm, in which all participating devices operate as equals without relying on a central coordinator. This architecture enhances system robustness and availability, eliminating single points of failure. Moreover, direct model exchanges among nodes can improve communication efficiency and strengthen privacy protection.
In \cite{shang2024energy}, this training architecture facilitates on-device SNN sharing and mutual knowledge transfer among Unmanned Aerial Vehicles (UAVs), thereby improving the real-time decision-making capabilities of the UAV swarm.
Similarly, IoT terminal devices collaborate to collect psychological data and perform depression detection using a spike-wise transformer model, enabling intelligent mental health monitoring deployed across diverse edge endpoints \cite{yang2025spike}.
This decentralized architecture is particularly well-suited for dynamic edge environments, where centralized coordination is infeasible or costly, and local autonomy and resilience are critical. 

\subsection{Security and Privacy for EdgeSNNs}
With the notable advancements of SNNs in academia and industry in recent years, numerous researchers have begun to investigate EdgeSNNs from a security perspective. In general, security research on SNNs focuses on two key aspects: privacy preservation and system robustness. The following subsections present a comprehensive literature review of existing studies that address these two dimensions.

\subsubsection{Privacy Risks and Countermeasures}
To enable secure and trustworthy on-device learning, deployable SNNs must protect both model confidentiality and the privacy of training data \cite{kundu2024recent}. In this context, two primary privacy threats emerge: \textbf{the leakage of locally collected data} and \textbf{the disclosure of model information}. These distinct forms of leakage correspond to different threat models, each necessitating dedicated privacy-preserving mechanisms.
PrivateSNN \cite{kim2022privatesnn} proposes an ANN-to-SNN training pipeline that utilizes synthetic data and employs a temporal-based learning rule to encrypt synaptic weights. This approach mitigates both data leakage and model information exposure, thereby enhancing the overall privacy of SNN deployment.
ADPSNN \cite{luo2025encrypted} introduces an adaptive differential privacy strategy that dynamically adjusts the privacy budget for gradient parameters based on the correlation between output spikes and target labels.
GASCNN \cite{khowaja2023spike} employs a generative adversarial network (GAN) to synthesize medical images from feature maps and trains a spike-wise ResNet using spike-based learning techniques. By encrypting model weights from the spatial domain into the temporal axis, this approach hinders reconstruction efforts and effectively mitigates both data and model leakage.
Encrypted-SNN \cite{luo2023encrypted} enhances model privacy by injecting noise into the gradients of both ANNs and SNNs during the ANN2SNN process, achieving improved privacy protection without sacrificing network performance. 
In \cite{li2023privacy,nikfam2023homomorphic}, homomorphic encryption is integrated with SNNs for an initial exploration, demonstrating promising prediction performance on encrypted data.

\subsubsection{Attacks and Countermeasures}

The deployment of SNNs on edge devices brings new security challenges, especially under adversarial threats during training. Compared to ANNs, SNNs rely on temporally dynamic learning, which increases the attack surface and complicates defense. 
These risks are further amplified by the decentralized and resource-constrained nature of edge environments, where traditional defenses are often inadequate.
As a result, improving the adversarial robustness of EdgeSNNs has become a key research focus, with growing efforts toward defense strategies tailored to the unique characteristics of SNNs.
In edge scenarios, adversaries may launch \textbf{training-time} and \textbf{inference-time} attacks to compromise the robustness of SNN-based services. 

For training-time attacks, the rate gradient approximation attack \cite{bu2023rate}, a method designed explicitly for SNNs, perturbs surrogate gradients and input spike activity to compromise the training process. Building on this, HART \cite{hao2024threaten} proposes a hybrid adversarial approach that dynamically manipulates both rate and temporal characteristics of spiking activity, effectively degrading training performance and offering insights into enhancing network robustness. A potential-dependent surrogate gradient \cite{lun2025towards} is introduced to establish a robust connection between the surrogate gradient and the SNN model, thereby improving the adaptability of adversarial attacks during training. 
Furthermore, dynamic backdoor attacks in neuromorphic data are first investigated in \cite{abad2022poster,abad2023sneaky}, demonstrating superior effectiveness compared to static and moving triggers, with high stealthiness. Notably, conventional image-domain defenses are shown to be ineffective against such attacks.

In terms of inference-time attacks, a novel adversarial example attack targeting the processing of raw event data from Dynamic Vision Sensors (DVS) is proposed in \cite{yao2024exploring}, revealing the vulnerability of SNNs to adversarial perturbations introduced during the transformation from raw event streams to grid-based representations at inference time. 
A comparison of visual information captured by neuromorphic and conventional visual sensors is depicted in Fig.~\ref{fig:cmp-imgs}.
Similarly, the sparse dynamic attack \cite{lun2025towards} generates targeted perturbations for sparse binary dynamic images by leveraging gradient information and finite differences, significantly reducing model accuracy with minimal modifications to the input.
In response to this security concern, a range of defense strategies have recently been proposed and actively explored.
HIRE-SNN \cite{kundu2021hire} introduces a spike-timing-dependent backpropagation mechanism aimed at improving the model's robustness against temporally distributed adversarial input noise. Similarly, SNN-RAT \cite{ding2022snn} performs a Lipschitz analysis of SNNs and introduces a regularized adversarial training scheme with low computational overhead. This method enhances model generalization across multiple adversarial $\epsilon$-neighborhoods. As a follow-up study, stochastic gating mechanisms are introduced in \cite{ding2024enhancing} as a biologically plausible regularizer during SNN training, mitigating error amplification under adversarial example attacks. In parallel, S-IBP and S-CROWN \cite{liang2022toward} establish output bounds for both spike-based and digital inputs in SNNs, enabling certified training to improve robustness against adversarial examples. 
Randomized smoothing coding \cite{mukhoty2024certified,wu2024rsc} is integrated with rate-coded inputs to provide provable robustness guarantees under input perturbations. In \cite{ding2024robust}, a training framework with modified neuronal dynamics is proposed to minimize mean-squared perturbations, thereby enhancing the robustness of SNNs against adversarial examples during evaluation. Additionally, FEEL-SNN \cite{xu2024feel} demonstrates the effectiveness of frequency encoding and an evolutionary membrane potential leakage factor in defending against various types of noise during model evaluation.

Security remains a critical yet underexplored dimension in the deployment of EdgeSNNs. Given the sensitivity of edge applications and the increasing prevalence of adversarial threats, ensuring the robustness of SNNs against attacks is of paramount importance. However, compared to their ANN counterparts, current attack strategies and defense mechanisms for SNNs remain relatively limited in scope and diversity. This disparity highlights a pressing need for systematic exploration of novel attack vectors and resilient training paradigms tailored to the unique temporal dynamics of SNNs. Addressing these gaps will be instrumental in enabling trustworthy and secure neuromorphic intelligence at the edge.

\section{Evaluation Methodology for EdgeSNNs}
\label{sec:eval}

As an emerging technology, neuromorphic chips have yet to converge on a standardized and commercially available hardware platform. As a result, much of the current research relies on algorithmic innovations deployed on general-purpose processors, such as GPUs and CPUs, which are not inherently optimized for neuromorphic workloads \cite{li2020spiking}. This hardware–algorithm disparity hinders the community’s ability to objectively evaluate novel approaches and identify promising directions for advancing efficiency, speed, and intelligence in EdgeSNNs. Therefore, establishing standardized and comprehensive evaluation protocols for SNNs on conventional hardware is essential. Such efforts not only enable fair comparisons across methods but also inform the co-evolution of algorithm design and hardware development.

\begin{figure*}[!t]
\centering
\includegraphics[width=0.8\textwidth]{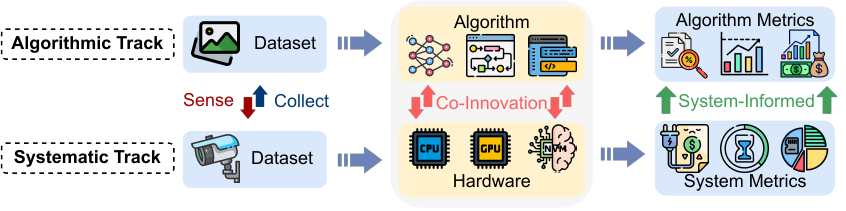}
\caption{Two tracks for evaluating EdgeSNNs. The best-performing results from each track can motivate future solutions for the other. In addition, system metrics and results can inform hardware-independent algorithmic complexity metrics.}
\label{fig:metrics-demo}
\end{figure*}

Notably, prior studies have made significant progress in establishing fair and rigorous benchmarking practices. For example, a comprehensive tool flow \cite{islam2024benchmarking} evaluates the energy efficiency and network sparsity of SNNs and their ANN counterparts across various high-level network architectures.
SNNBench \cite{tang2023snnbench} presents an end-to-end benchmark for evaluating the accuracy of SNNs, encompassing both training and inference on CPUs and GPUs, while accounting for spiking neuron dynamics and various learning paradigms.
PRONTO \cite{matinizadeh2024open} introduces a unified framework for verifying and benchmarking neuromorphic chip designs in terms of inference accuracy, area, power consumption, and throughput, using QUANTISENC \cite{matinizadeh2024fully}, a software-defined SNN hardware platform.
STEP \cite{shen2025step} proposes a unified benchmarking framework for Spiking Transformers that supports a broad range of tasks, including classification, segmentation, and detection. It also introduces a unified analytical model for energy estimation, taking into account spike sparsity, bitwidth, and memory access.
Neuromorphic Sequential Arena \cite{chen2025neuromorphic} provides a comprehensive, application-oriented evaluation benchmark for real-world temporal processing tasks utilizing SNNs, assessing task performance, training speed, memory usage, and energy efficiency.
SpikeSim \cite{moitra2023spikesim} enables hardware-realistic evaluation of SNNs in terms of performance, energy consumption, latency, and area, when mapped onto a practical monolithic compute-in-memory architecture.
In \cite{parker2022benchmarking}, an SNN with two densely connected layers is implemented on both Loihi \cite{davies2018loihi} and a custom CPU to compare power consumption, energy usage, and execution time. 

Overall, existing evaluation benchmarks exhibit notable limitations when applied to EdgeSNNs. Most tools lack system-level modeling that accounts for software stack overhead and real-world I/O latency, and they fail to capture runtime dynamics across heterogeneous edge platforms. Moreover, these benchmarks primarily focus on static classification tasks, often neglecting real-time responsiveness, cross-task generalization, and training–deployment mismatches that are prevalent in edge scenarios. Such gaps hinder a comprehensive understanding of SNN efficiency and practical applicability in resource-constrained, latency-sensitive environments.

To tackle these limitations, we adopt the NeuroBench guidelines \cite{yik2025neurobench} and propose two complementary evaluation tracks to facilitate fair and rigorous assessment of EdgeSNN applications: (1) the \textbf{Algorithmic Track}, which focuses on hardware-independent performance benchmarking, and (2) the \textbf{Systematic Track}, which evaluates fully deployed solutions across diverse hardware platforms. As illustrated in Fig.~\ref{fig:metrics-demo}, these two tracks jointly support the agile co-development of algorithms and systems for EdgeSNNs.

\subsection{Algorithmic Track} 

The algorithmic track defines platform-agnostic primary metrics that are broadly applicable to diverse EdgeSNN solutions. These include: (1) \textbf{Correctness} metrics, which evaluate the quality of model predictions for specific tasks, such as \textit{accuracy}, \textit{mean average precision} (mAP), and \textit{mean squared error} (MSE), and (2) \textbf{Complexity} metrics, which assess the theoretical computational demands of the algorithm. By assuming a digital, time-stepped execution of SNN-based algorithms, typical complexity metrics include:
\begin{itemize}
    \item \textbf{Footprint}: The theoretical memory (in bytes) required to represent a model. This metric provides a comprehensive summary of memory demands and can be further decomposed into components like synaptic weight count, trainable neuronal parameters, and data buffer requirements.
    \item \textbf{Sparsity}: There are two types of sparsity in SNN-based models. \textit{Connection sparsity} refers to the ratio of zero weights to total weights across all layers, ranging from 0 (fully connected) to 1 (no connections). It reflects structural pruning or inherently sparse architectures \cite{chen2022state}. \textit{Activation sparsity} measures the average proportion of inactive neuron outputs across all neurons, layers, time steps, and test samples, where 0 indicates fully active neurons and 1 denotes completely silent ones \cite{shi2024towards}.
    \item \textbf{Synaptic Operations}: This metric represents the average number of synaptic operations (SOPs) per model execution, determined by neuron activations and their corresponding fan-out synapses. SOPs are categorized into dense SOPs, multiply-accumulate operations (MACs), and accumulate operations (ACs). Dense SOPs include all operations, regardless of activation or weight sparsity, reflecting computation on hardware without sparsity support. In contrast, MACs and ACs account only for non-zero activations and weights, offering a more accurate estimate of computation on sparsity-aware hardware (e.g., neuromorphic chips). Specifically, SOPs involving non-binary activations are counted as MACs, while those with spikes are treated as ACs.
\end{itemize}
In particular, footprint and connection sparsity are static metrics that can be analytically derived solely from the model architecture. In contrast, activation sparsity, SOPs, and correctness are workload-dependent metrics that require model execution or simulation on specific tasks and platforms for accurate evaluation.

\subsection{Systematic Track}
The systematic track evaluates the execution time, throughput, and efficiency of systems running algorithms optimized for specific hardware platforms. It focuses on fully deployed systems at the task level, emphasizing overall system performance rather than individual components. This enables direct comparison based on problem-solving capabilities and provides an objective measure of solution efficiency. 
Task-specific metrics are employed to evaluate the EdgeSNN systems under test (SUTs), including:

\begin{itemize} 
    \item \textbf{Timing}: Timing performance encompasses sample \textit{throughput} and execution time (i.e., \textit{latency}). Throughput typically corresponds to batched, offline inference, whereas latency reflects a streaming scenario in which each inference starts only after the previous one has finished. For tasks that require the simultaneous completion of multiple requests within a fixed time window, both metrics should be reported.
    \item \textbf{Efficiency}: As energy efficiency is central to both biological systems and neuromorphic research, \textit{energy consumption} must be included in system metrics despite measurement challenges. Similar to timing performance, efficiency metrics should be benchmark-specific—for example, average power for always-on tasks, and peak power or energy cost per inference for high-throughput workloads.
    \item \textbf{Resilience}: Quantifying system resilience and robustness is essential when benchmarking SUTs against disruptions such as communication failures \cite{borsos2022resilience}, resource reallocations \cite{zhang2024spiking}, and hardware faults \cite{putra2021sparkxd}. Key metrics include the Recovery Time Objective (RTO), which specifies the maximum tolerable downtime; the Mean Time to Repair (MTTR), which measures recovery speed; and the Mean Time Between Failures (MTBF), which reflects operational stability \cite{schuman2020resilience}. Together, these metrics establish a cohesive foundation for evaluating and enhancing system dependability.

\end{itemize}

Ideally, standardized measurement methodologies, covering power tools, interfaces, data loading, and software, are essential for a fair evaluation of EdgeSNNs. However, the diversity in implementation, hardware integration, and development maturity poses significant challenges to consistent benchmarking \cite{roy2019towards,basu2022spiking}.
Hence, as a first step toward consistently evaluating EdgeSNNs, we advocate for clear documentation guidelines as a foundation for a shared methodology across diverse platforms.
For instance, in the absence of practical tools for measuring computational energy, numerous studies \cite{lv2024efficient, ijcai2024p596, yao2023attention, shi2024towards, zheng2023neuroradar} estimate efficiency by combining effective SOPs with energy parameters of specific chips. Despite variations in measurement, providing key details enables meaningful comparison and supports future consistency through transparency.

\section{Open Challenges and Future Directions}
\label{sec:challenge}

\begin{figure}[!t]
    \centering
    \includegraphics[width=\columnwidth]{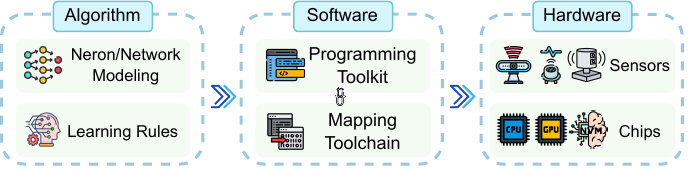}
    \caption{Workflow for developing an EdgeSNN system: software serves as a critical bridge, linking algorithmic implementation via programming toolkits to hardware deployment through mapping toolchains.}
    \label{fig:develop-edgesnn}
\end{figure}

Although EdgeSNNs have shown promising performance, the following sections will discuss their challenges and potential future research directions.
As demonstrated in Fig.~\ref{fig:develop-edgesnn}, the development of an EdgeSNN system necessitates a holistic algorithm–software–hardware co-design paradigm, with each layer presenting distinct challenges and research opportunities.

\subsection{Heterogeneous Hardware in EdgeSNNs}
In edge deployment, SNN-based systems implemented on heterogeneous hardware platforms—including digital accelerators, neuromorphic cores, FPGAs, and mixed-signal circuits—aim to balance strict constraints on power, latency, and bandwidth. However, integrating these diverse processing elements presents several critical challenges.
First, discrepancies in Instruction Set Architecture (ISA), Application Binary Interface (ABI), and Application Programming Interface (API) across compute units hinder software portability and complicate dynamic workload scheduling \cite{seekings2024towards}.
Second, mapping SNNs onto multi-core neuromorphic fabrics requires careful partitioning to minimize Network-on-Chip (NoC) latency and energy consumption, as suboptimal clustering can increase spike communication delays and power usage \cite{xiao2022optimal}.
Third, while enabling intelligent models with neural dynamics at the edge is an emerging direction, hardware support for sparse, event-driven computation remains inconsistent across platforms, limiting the scalability of SNNs’ intrinsic mechanisms \cite{liu2024activen}.
Finally, although memristive mixed-signal designs offer promising scalability for synaptic integration, they introduce variability and calibration challenges, necessitating standardization for reliable deployment.

To surmount these barriers, future work must pursue unified heterogeneous abstraction layers to harmonize interfaces across platforms; develop mapping-aware communication protocols that adaptively co-locate spike clusters to reduce NoC load; co-optimize sparse-aware circuit architectures supporting event compression and low-power mixed-signal memory; and deliver comprehensive co‑design tools and benchmarks linking device-level structures (e.g., FPGAs, memristors, neuromorphic cores) up to system-level compilers, schedulers, and evaluation suites.
Addressing these interwoven research dimensions—from abstraction to tooling—will be essential to the practical realization of real-time, adaptive, and energy-efficient EdgeSNN systems on heterogeneous platforms.

\subsection{Algorithm-Software-Hardware Co-Design in EdgeSNNs} 
In the holistic co-design of EdgeSNNs, software bridges the algorithmic and hardware layers. Yet, significant open challenges remain in two critical dimensions: programming toolkits for algorithm authorship and mapping toolchains for hardware deployment. 

On the algorithm–software front, existing SNN-oriented programming toolkits—summarized in Table~\ref{tab:overview-toolkit}—are still in their early stages and face several key challenges in supporting edge deployment.
First, sparse-event modeling is not natively supported. Most mainstream toolkits are built upon dense tensor algebra, whereas spiking networks operate on asynchronous and inherently sparse spike streams. Although toolkits such as SpikingJelly \cite{fang2023spikingjelly} and Slax \cite{summe2025slax} provide PyTorch- or JAX-style programming interfaces for SNNs, they largely lack explicit support for event-driven constructs such as spike histograms, inter-spike interval tracking, or hardware resource estimation based on event rates \cite{xue2023edgemap}. Without spike-aware static profilers and compile-time estimators, developers are unable to predict whether a given model will exceed memory, bandwidth, or power budgets on edge devices.
Secondly, resource-aware quantization and sparsification remain underexplored. Unlike ANN frameworks that offer well-integrated support for quantization, pruning, and compression under hardware constraints, current EdgeSNN toolkits lack streamlined mechanisms for jointly optimizing spiking thresholds, weight precision, and sparsity. There are no unified APIs for quantization strategies tailored to spike dynamics, nor tools to analyze trade-offs between sparsity, quantization, and accuracy on resource-constrained hardware \cite{carpegna2024spiker+}.
Transitioning to software–hardware mapping, the cross-platform portability of existing toolchains remains limited. Most toolchains are tightly coupled to specific runtimes—such as Loihi, FPGAs, or CPUs—due to hard-coded hardware backends and the absence of a unified intermediate representation (IR) layer analogous to ONNX. This rigidity significantly restricts model reuse: a network optimized for one edge platform (e.g., Loihi) cannot be easily re-targeted to another without re-engineering the entire toolchain or compilation pipeline.

\begin{table}[!t]
\centering
\caption{Overview of SNN-oriented open-source programming toolkits.}
\label{tab:overview-toolkit}
\resizebox{\columnwidth}{!}{%
\begin{threeparttable}
\begin{tabular}{lrrr}
\toprule
\rowcolor[HTML]{f3fbf2} 
\textbf{Library} & \textbf{Affiliation} & \textbf{Year} & \textbf{Language}  \\ 
\midrule \midrule
PyNCS \cite{stefanini2014pyncs} & University of Heidelberg & 2014 & Python \\
BindsNet \cite{hazan2018bindsnet} & University of Massachusetts Amherst & 2018 & Python \\
cuSNN \cite{paredes2019unsupervised} & Delft University of Technology & 2018 & C++ \\
PySNN$^\dagger$ & Open Community & 2019 & Python \\
SpykeTorch \cite{mozafari2019spyketorch} & CNRS & 2019 & Python \\
Sinabs \cite{liu2019live} & SynSense & 2019 & Python \\
SpyTorch$^\ddagger$ & University of Basel & 2019 & Python \\
SpikingJelly \cite{fang2023spikingjelly} & Peking University & 2020 & Python \\ 
Norse \cite{pehle2021norse} & University of Heidelberg & 2021 & Python \\ 
SNNtorch \cite{eshr2023train} & University of Michigan & 2021 & Python \\
Spaic \cite{hong2024spaic} & Zhejiang University & 2022 & Python \\ 
Spyx \cite{heckel2024spyx} & University of Cambridge & 2023 & JAX  \\
SNNAX \cite{lohoff2024snnax} & PGI 15 & 2024 & JAX  \\
Slax \cite{summe2025slax} & University of Notre Dame & 2025 & JAX \\
 \bottomrule
\end{tabular}%
\begin{tablenotes}
\item[$\dagger$] https://github.com/BasBuller/PySNN
\item[$\ddagger$] https://github.com/fzenke/spytorch
\end{tablenotes}
\end{threeparttable}
}
\end{table}

To advance EdgeSNN usability and performance, future research directions to address these software-level gaps include: 
(1) Designing compiler infrastructures capable of parsing high-level SNN descriptions and statically estimating event loads, memory pressure, and energy overheads, thereby enabling developers to optimize model structures preemptively before deployment. 
(2) Embedding event-driven quantization, pruning, and threshold optimization into toolkits, supported by auto-tuning mechanisms that profile accuracy against edge resource constraints (e.g., energy, memory) and automatically adjust network hyperparameters.
(3) Developing SNN-specific IRs, analogous to ONNX but incorporating constructs for spatiotemporal dynamics and spike semantics, to enable compilation across multiple hardware backends and achieve true hardware portability.
(4) Embedding lightweight runtime monitors within toolkits to measure on-device event traffic, buffer utilization, and latency, enabling dynamic tuning (e.g., lowering spike thresholds or pruning inactive pathways) without requiring complete remapping.
(5) Programming toolkits should seamlessly integrate with lightweight mapping engines to generate annotated SNN graphs enriched with metadata—such as core-cost estimates and locality hints—that guide the mapping toolchain. 
By enabling high-level programmability that is both spike-aware and resource-sensitive, and coupling it with a portable backend targeting strategy, future EdgeSNN toolkits can provide developer-level experience and deployment predictability comparable to mainstream deep learning frameworks, while respecting the sparse, temporal, and heterogeneous characteristics of edge neuromorphic systems.

\subsection{Collected Data Limitations in EdgeSNNs} 
In resource-constrained IoT environments, EdgeSNNs face significant limitations due to severe data scarcity \cite{gong2024delta} and distribution drift \cite{diao2024oebench}, which arise from privacy restrictions, sporadic sampling rates, and intermittent connectivity. 
These conditions compel devices to collect limited, imbalanced, and fragmented datasets, which hamper the SNN’s ability to learn rich temporal dynamics, often resulting in underfitting and diminished generalization \cite{fahy2022scarcity}.
Subsequently, dynamic environmental changes—manifested in shifting sensor characteristics or evolving user behaviors—induce distribution drift that gradually degrades model performance. While SNNs naturally support temporal integration and incremental plasticity through mechanisms such as STDP, existing EdgeSNN frameworks lack lightweight and reliable drift-detection mechanisms tailored to the constraints of edge devices.
Furthermore, although recent work in brain-inspired online adaptation shows promise \cite{zhu2024online,jiang2024ndot}, these methods are yet to be validated under severe data scarcity or inconsistent streaming conditions typical of IoT deployments. This contrast underscores a critical gap: while adaptation algorithms exist, they often presuppose continuous data flows and the absence of privacy-driven sparsity, leaving the challenge of robust learning under both labeled-data scarcity and abrupt distribution shifts largely unaddressed.

Future research should thus focus on developing data augmentation and few-shot learning methods to alleviate label scarcity while preserving spiking temporal features \cite{huo2023bio}, designing edge-compatible drift detection ensembles to identify distribution changes with minimal overhead timely \cite{bodyanskiy2024ensemble}, implementing incremental learning strategies to enable lifelong adaptation without catastrophic forgetting \cite{zheng2024continuous}, and establishing holistic evaluation metrics that jointly consider accuracy, adaptation latency, energy efficiency, and memory footprint \cite{yu2025eccsnn}. Collectively, these advances will empower EdgeSNNs to operate robustly and efficiently in the evolving and sparse data landscapes characteristic of IoT systems.

\subsection{Open-Source Benchmarking Platforms in EdgeSNNs} 
Current SNN-related open-source benchmarking platforms \cite{tang2023snnbench,shen2025step} primarily emphasize model accuracy and end-to-end training or inference workloads, making them valuable for algorithmic comparisons. However, they fall short in evaluating practical key performance indicators (KPIs) critical for real-world deployment of EdgeSNNs, such as energy consumption, latency, throughput, scalability, and adaptability under resource constraints. Likewise, existing neuromorphic frameworks \cite{ostrau2022benchmarking,ostrau2020benchmarking} focus on functional correctness and performance consistency, but lack support for assessing communication overhead, sparse event rates, and real-time responsiveness.
This imbalance poses several challenges. On one hand, benchmark suites often rely on unrealistic CPU/GPU-centric workloads that overlook the sporadic event traffic and constrained compute budgets of neuromorphic accelerators. On the other hand, many frameworks provide siloed metrics that fail to capture the trade-offs among energy, latency, and hardware utilization. While general-purpose benchmarks such as BenchIP \cite{tao2018b} support multi-dimensional metrics (e.g., operations per joule, energy-delay product) \cite{islam2024benchmarking,aymone2024benchmarking}, they are not specifically designed to accommodate SNN-specific properties like spike sparsity, temporal precision, and the architectural idiosyncrasies of neuromorphic hardware.

To overcome these shortcomings, future research should focus on building EdgeSNN-specific benchmarking frameworks that integrate the following key capabilities:
\begin{enumerate}
    \item \textbf{Multi-dimensional KPI instrumentation}, encompassing per-spike energy consumption, network latency under real-world duty-cycle workloads, throughput under bursty event arrivals, and scalability across multiple low-power neuromorphic cores.
    \item \textbf{Configurable workload generators} that simulate realistic IoT and edge event streams—characterized by intermittent activity, spatial clustering, and temporal sparsity—as commonly observed in neuromorphic edge sensor data.
    \item \textbf{Standardized interfaces and metadata formats} to harmonize model descriptions, hardware backends, and benchmark outputs—potentially via extensions to existing frameworks like emerging SNN-specific IRs.
    \item \textbf{Adaptive evaluation flows} that capture dynamic performance under schema evolution and resource variability, thereby enabling meaningful and comparable assessments across heterogeneous platforms.
    \item \textbf{Reproducible, open-source benchmarking toolkits} developed as community resources—featuring Dockerized workloads, plug-and-play hardware adapters, and comprehensive KPI dashboards—designed to support both algorithm-software–hardware co-design and systematic hardware evaluation. 
\end{enumerate}
By embracing a holistic evaluation approach—measuring models not only for accuracy, but also for energy efficiency, latency under event-driven regimes, and behavior under real-world edge conditions—such benchmarking platforms can catalyze advances in both EdgeSNN algorithms and hardware design, propel standardization, and accelerate adoption in practical IoT applications.

\section{Conclusion}
\label{sec:conclude}

In this survey, we presented a comprehensive review of research on the foundations of EdgeSNNs, encompassing neuronal and network modeling, learning algorithm design, and hardware platform support. We further outlined key practical considerations for EdgeSNN, including edge deployment and inference, on-device training and updating, as well as associated security and privacy concerns. From each perspective, we systematically categorized existing and emerging methods, summarized representative architectures, and highlighted the strategies adopted across various approaches. In addition, we proposed a fair and rigorous evaluation methodology for EdgeSNNs that spans both algorithmic and system-level dimensions. Finally, we discussed open challenges in this domain and identified promising future research directions.
Given the growing demand for intelligent, low-power, and low-latency edge AI solutions, EdgeSNNs are poised to play a transformative role in next-generation computing paradigms. Their unique event-driven, sparse, and bio-inspired processing characteristics make them particularly suitable for addressing the stringent energy and responsiveness requirements of edge environments. As advances in neuromorphic hardware, learning algorithms, and system-level integration continue to mature, we anticipate that EdgeSNNs will enable a new wave of ubiquitous and energy-efficient cognitive services across diverse domains, such as autonomous vehicles, wearable devices, intelligent sensing, and industrial IoT. We hope this survey will serve as a valuable resource for researchers seeking to deepen their understanding of EdgeSNNs and to foster continued innovation in this rapidly evolving field.

\section*{Acknowledgments}
The work of this paper is supported by the National Key Research and Development Program of China under Grant 2022YFB4500100, the National Natural Science Foundation of China under Grant 62125206, and the Zhejiang Provincial Natural Science Foundation of China under Grant No. LD24F020014, the National Key Research and Development Program of China No. 2024YDLN0005, and the Regional Innovation and Development Joint Fund of the National Natural Science Foundation of China No. U22A6001.


\bibliographystyle{IEEEtran}
\bibliography{IEEEabrv}

\begin{IEEEbiography}[{\includegraphics[width=1in,height=1.25in, clip,keepaspectratio]{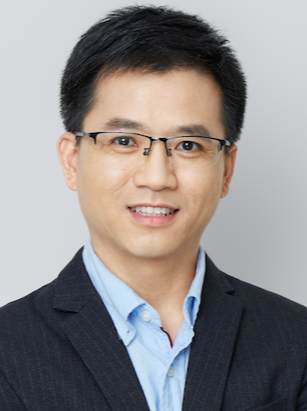}}]
{Shuiguang Deng} (Senior Member, IEEE) received the BS and PhD degrees in computer science from Zhejiang University, China, in 2002 and 2007, respectively. He is currently a full professor in the College of Computer Science and Technology at Zhejiang University. He previously worked as a visiting scholar at the Massachusetts Institute of Technology in 2014 and at Stanford University in 2015. His research interests include edge computing, service computing, cloud computing, and business process management. He serves as an associate editor for the journals IEEE Transactions on Services Computing, Knowledge and Information Systems, Computing, and IET Cyber-Physical Systems: Theory and Applications. To date, he has published over 100 papers in peer-reviewed journals and conferences. In 2018, he received the Rising Star Award from IEEE TCSVC. He is a fellow of the IET.
\end{IEEEbiography}

\begin{IEEEbiography}[{\includegraphics[width=1in,height=1.25in, clip,keepaspectratio]{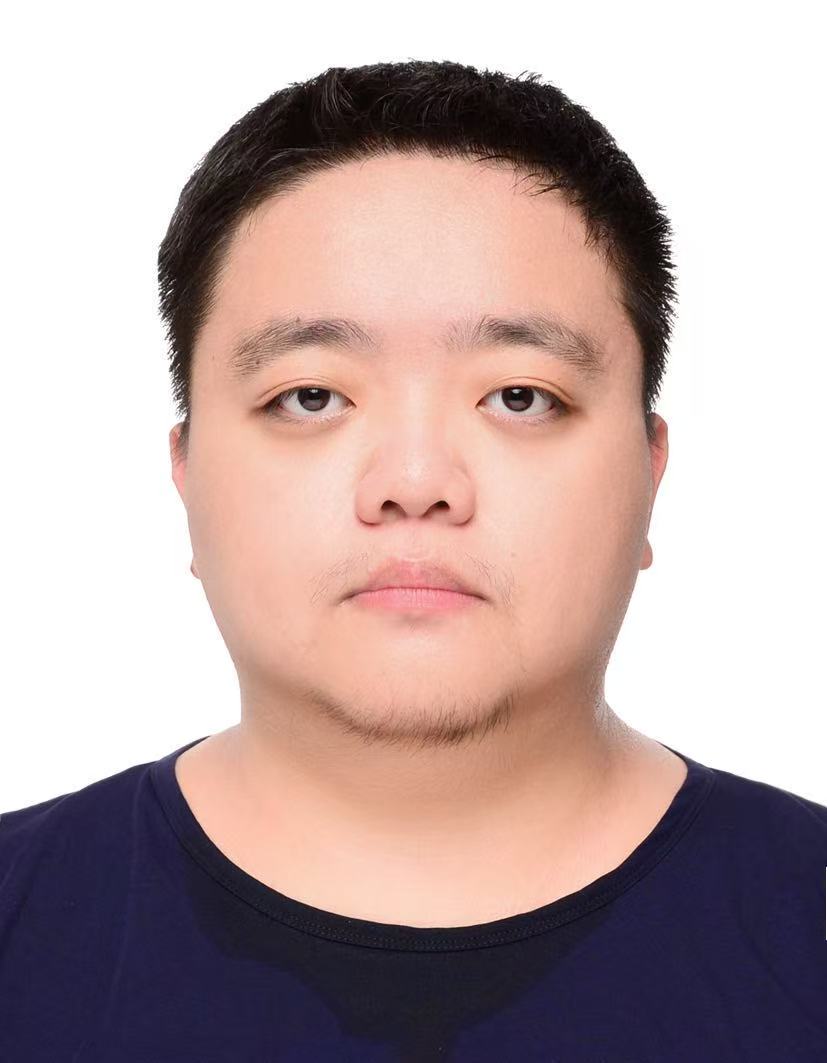}}]
{Di Yu} received the B.E. in computer science and M.M. in management science degrees from the Shanghai University of Finance and Economics, China, in 2017 and 2019, respectively, and the M.Sc. degree in IT Business from Singapore Management University, Singapore, in 2023. He is currently pursuing a Ph.D. degree in the School of Computer Science and Technology at Zhejiang University, China. His current research interests include brain-inspired computing, edge intelligence, data science, and recommendation systems. He has published several papers in top conferences and journals, including IJCAI, TSC, and ICML.
\end{IEEEbiography}

\begin{IEEEbiography}[{\includegraphics[width=1in,height=1.25in, clip,keepaspectratio]{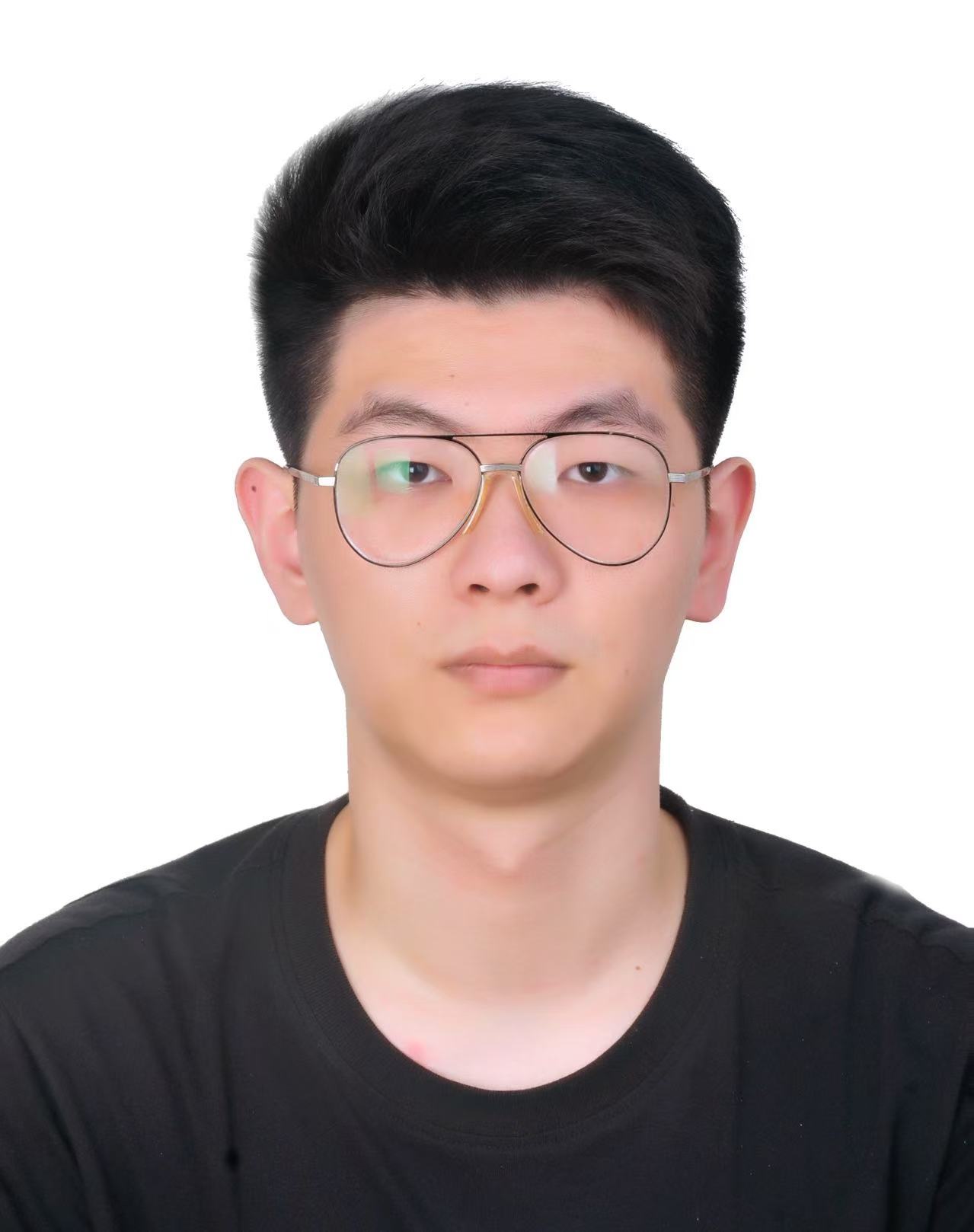}}]
{Changze Lv} received the B.E. degree in software engineering from Fudan University, China, in 2022. He is currently pursuing the Ph.D. degree with the School of Computer Science, Fudan University, China. His current research interests include brain-inspired computing, large language models, and time-series analysis. He has published several papers in top conferences and journals, including ICML, NeurIPS, and ICLR.
\end{IEEEbiography}

\begin{IEEEbiography}[{\includegraphics[width=1in,height=1.25in, clip,keepaspectratio]{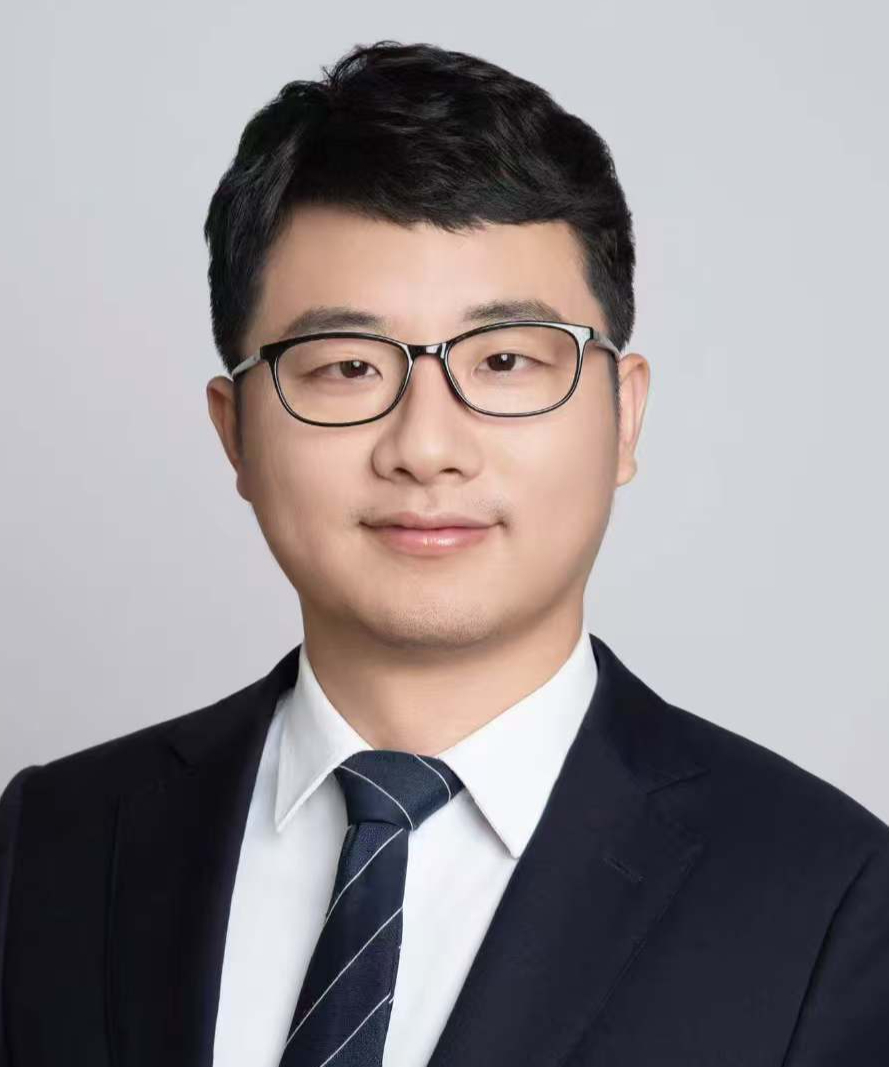}}]
{Xin Du} is an Assistant Professor at the School of Software Technology, Zhejiang University, China. He received a Ph.D. in computer science from Fudan University in 2024. His research interests include service computing, distributed systems, and brain-inspired computing. He has published several papers in flagship conferences and journals, including IEEE ICWS and the IEEE Transactions on Parallel and Distributed Systems. He has received the Best Student Paper Award of IEEE ICWS 2020 and IEEE ICWS 2023.
\end{IEEEbiography}

\begin{IEEEbiography}[{\includegraphics[width=1in,height=1.25in, clip,keepaspectratio]{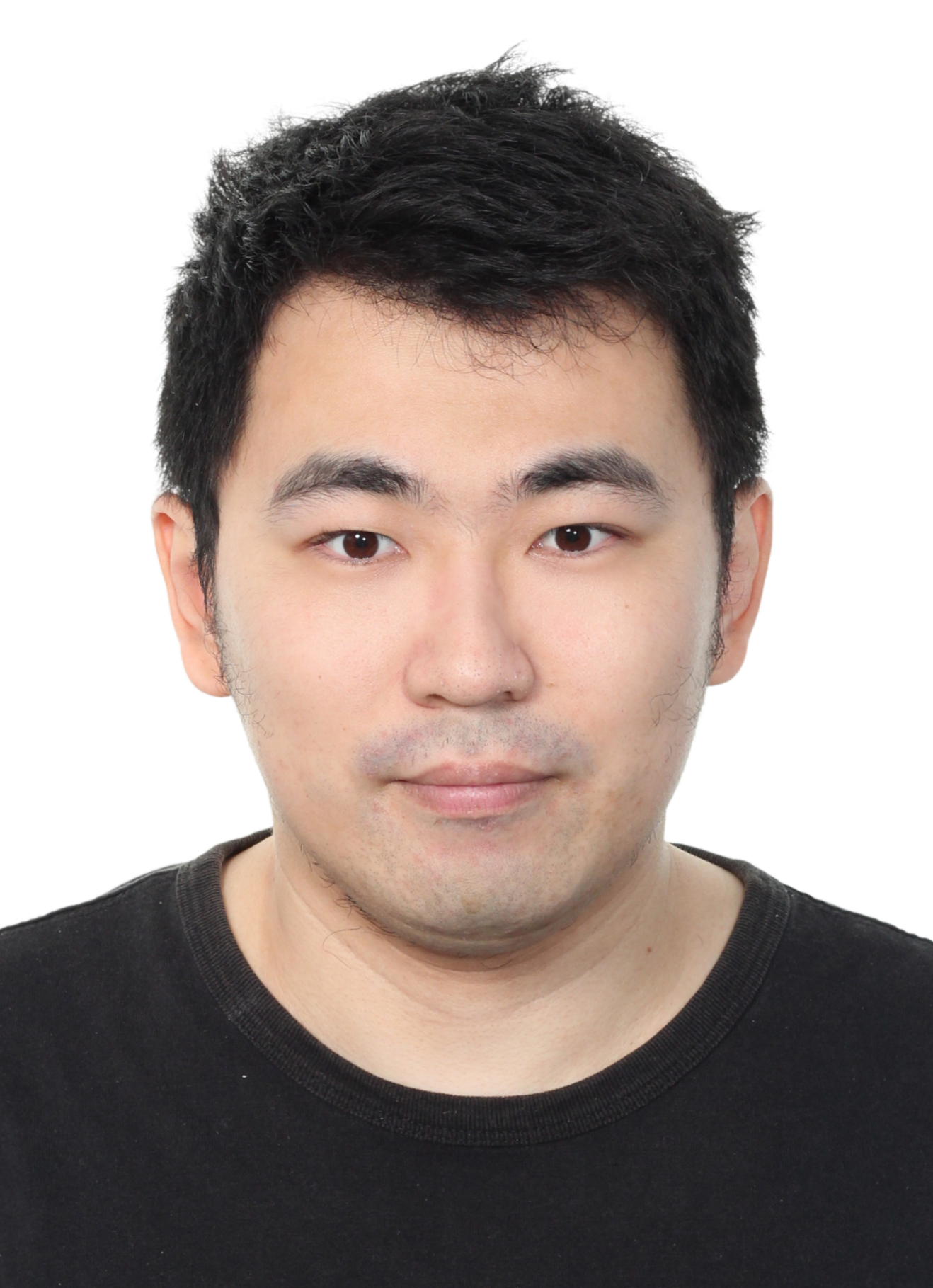}}]
{Linshan Jiang} is currently a Research Fellow at the Institute of Data Science, National University of Singapore. He obtained his Ph.D. degree in Computer Science and Engineering from Nanyang Technological University, Singapore, in 2022. He has published several papers in the top conferences and journals in CPS-IoT. His research interests focus on privacy and security in distributed AI systems, including federated and collaborative learning, blockchain-enabled AI, and resilient AIoT systems.
\end{IEEEbiography}

\begin{IEEEbiography}[{\includegraphics[width=1in,height=1.25in, clip,keepaspectratio]{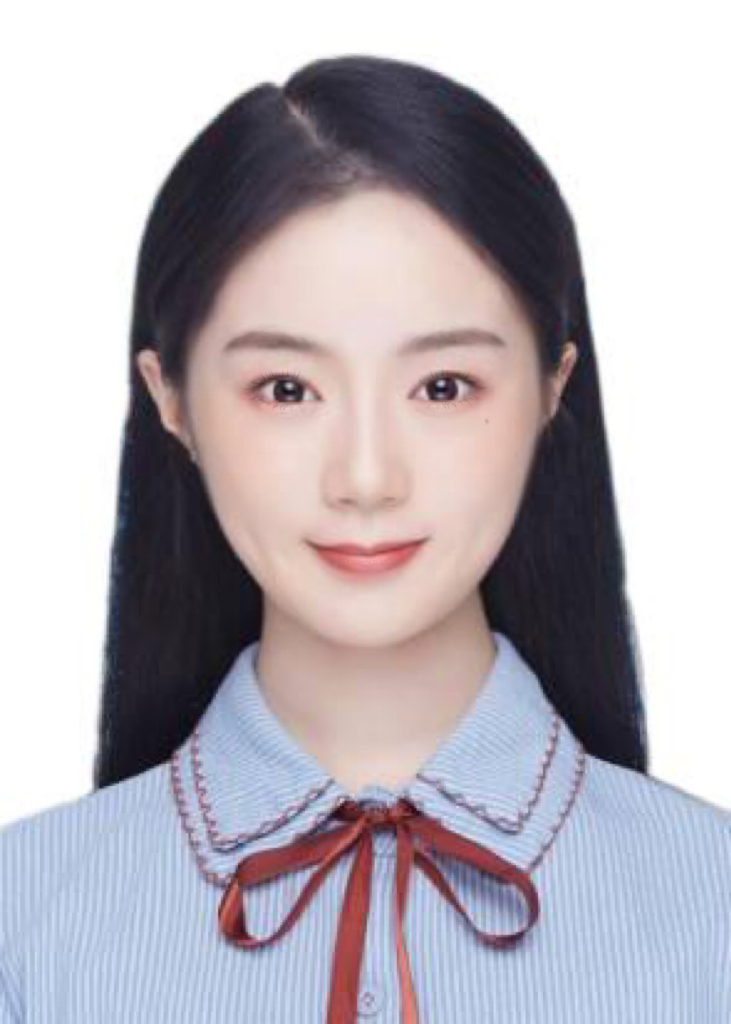}}]
{Xiaofan Zhao} received the B.Sc. degree in Computer Science and Technology from Central China Normal University, Wuhan, China, in 2020. She is currently pursuing a Ph.D. degree in the College of Computer Science and Technology at Zhejiang University, Hangzhou, China. Her research interests include service computing, brain-inspired computing, and edge intelligence.
\end{IEEEbiography}

\begin{IEEEbiography}[{\includegraphics[width=1in,height=1.25in, clip,keepaspectratio]{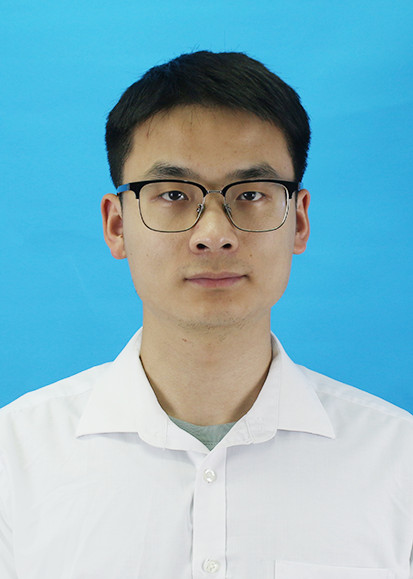}}]
{Wentao Tong} received the B.E. degree from Zhejiang University of Technology, Hangzhou, China, in 2023. He is currently pursuing a PhD degree in brain-inspired computing at the School of Computer Science, Zhejiang University. His research interests include brain-inspired computing, compiler design, edge computing, and service computing.  
\end{IEEEbiography}

\begin{IEEEbiography}[{\includegraphics[width=1in,height=1.25in,clip,keepaspectratio]{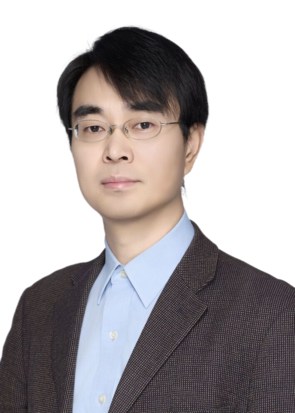}}]
{Xiaoqing Zheng} is an Associate Professor of the School of Computer Science, Fudan University, Shanghai, China. He received his Ph.D. degree in Computer Science from Zhejiang University in 2007. He was an international faculty fellow at the Sloan School of Management at the Massachusetts Institute of Technology (MIT). He also visited the natural language processing and machine learning groups at the University of California, Los Angeles (UCLA), as a visiting researcher from 2019 to 2020. His research interests include Natural Language Processing, Management Science, and Machine Learning.
\end{IEEEbiography}

\begin{IEEEbiography}[{\includegraphics[width=1in,height=1.25in,clip,keepaspectratio]{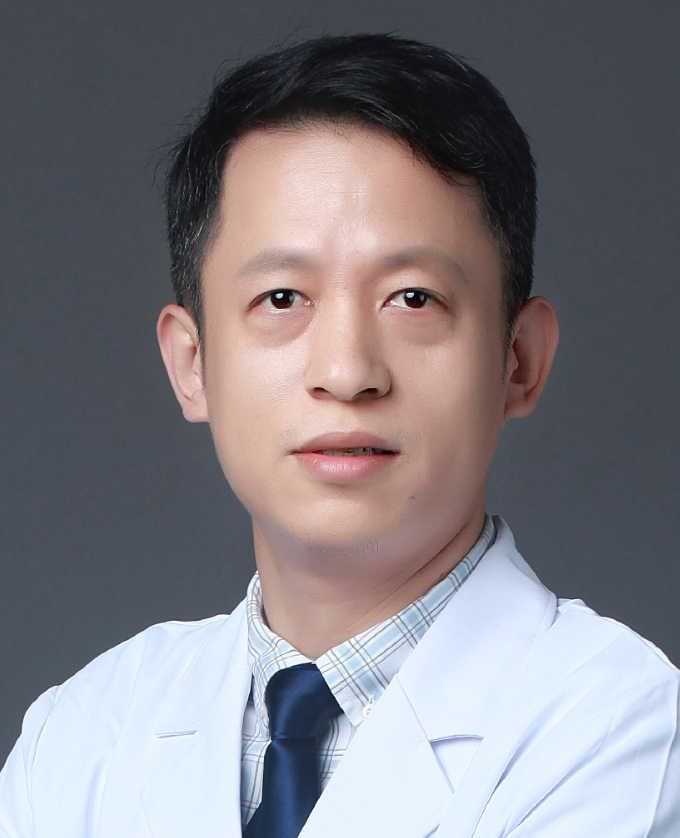}}]
{Weijia Fang} Since Aug 1998, he has completed both his rotating internship and residency at the First Affiliated Hospital, Zhejiang University School of Medicine, and was subsequently awarded clinical fellowships in medical oncology at FAHZU. He has accumulated broad clinical experience in the chemotherapy and biotherapy of solid malignancies, mainly in digestive system cancer. He has participated as PI in more than 20 international and domestic clinical trials involving anticancer therapies. He is a full member of ASCO and the Chinese Society of Clinical Oncology(CSCO). He has been an author and co-author of several original research publications in national and international peer-reviewed scientific and medical journals.
\end{IEEEbiography}

\begin{IEEEbiography}[{\includegraphics[width=1in,height=1.25in,clip,keepaspectratio]{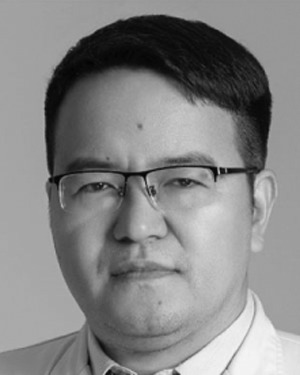}}]
{Peng Zhao} received the PhD degree in oncology from Sun Yat-sen University, Guangzhou, China, in 2007. He works with the Department of Medical Oncology, the First Affiliated Hospital, Zhejiang University School of Medicine, Hangzhou, China. He is currently leading some research projects supported by the National Natural Science Foundation of China. His research interests include artificial intelligence and edge computing in medical oncology.
\end{IEEEbiography}

\begin{IEEEbiography}[{\includegraphics[width=1in,height=1.25in,clip,keepaspectratio]{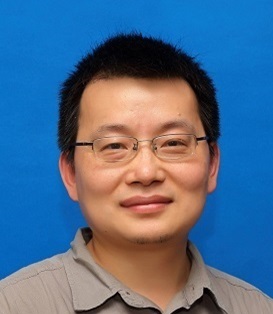}}]
{Gang Pan} (Senior Member, IEEE) is a distinguished professor in the College of Computer Science and Technology at Zhejiang University, where he also serves as the Director of the State Key Laboratory of Brain-Machine Intelligence. He earned his B. Eng. and Ph.D. degrees from Zhejiang University in 1998 and 2004, respectively. His research interests include brain-machine interfaces, brain-inspired computing, artificial intelligence, and pervasive computing. He has received numerous honors, including the NSFC Distinguished Young Scholars, the IEEE TCSC Award for Excellence (Middle Career Researcher), and the CCF-IEEE CS Young Scientist Award. Additionally, he has been recognized with the National Science and Technology Progress Award, two Test-of-Time paper awards, and multiple Best Paper awards. He serves as an associate editor for multiple prestigious journals, including IEEE Transactions on Neural Networks and Learning Systems.
\end{IEEEbiography}

\begin{IEEEbiography}
[{\includegraphics[width=1in,height=1.25in,clip,keepaspectratio]{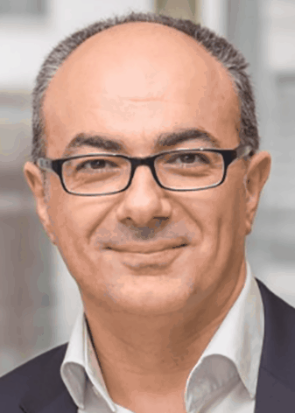}}]
{Schahram Dustdar} (Fellow, IEEE) is currently a Full Professor of computer science (informatics) heading the Distributed Systems Group at the Technische Universität Wien (TU Wien), Vienna, Austria. He is also a research Professor with ICERA Barcelona, Spain. Dr. Dustdar is an elected member of the Academia Europaea, the Academy of Europe, where he has served as Chairperson of the Informatics Section for multiple years. He has served as the President of the Asia-Pacific Artificial Intelligence Association (AAIA) from 2021 till 2023 and as a Fellow since 2021. He has been an EAI Fellow since 2021 and an I2CICC Fellow since the same year. He was a member of the IEEE Computer Society Fellow Evaluating Committee from 2022 to 2025. He was a recipient of multiple awards, including the TCI Distinguished Service Award in 2021, the IEEE Technical Community on Services Computing (TCSVC) Outstanding Leadership Award in 2018, the IEEE TCSC Award for Excellence in Scalable Computing in 2019, the ACM Distinguished Scientist in 2009, the ACM Distinguished Speaker in 2021, and the IBM Faculty Award in 2012. He was the Founding Co-Editor-in-Chief of ACM Transactions on Internet of Things (ACM TIoT). He is the Editor-in-Chief of Computing (Springer). He is an Associate Editor of IEEE TRANSACTIONS ON SERVICES COMPUTING, IEEE TRANSACTIONS ON CLOUD COMPUTING, ACM Computing Surveys, ACM Transactions on the Web, and ACM Transactions on Internet Technology and on the Editorial Board of IEEE INTERNET COMPUTING and the IEEE Computer Society.
\end{IEEEbiography}

\begin{IEEEbiography}
[{\includegraphics[width=1in,height=1.25in,clip,keepaspectratio]{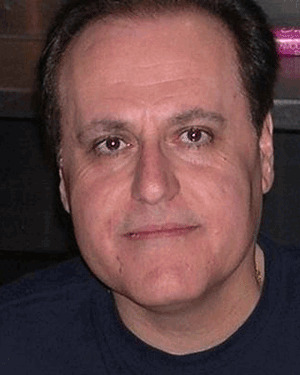}}]
{Albert Y. Zomaya} (Fellow, IEEE) is currently the Peter Nicol Russell Chair Professor of computer science and the Director of the Centre for Distributed and High-Performance Computing, The University of Sydney, Sydney, NSW, Australia. He has published more than 800 scientific papers and articles and is a (co)author/editor of more than 30 books. As a sought-after speaker, he has delivered more than 300 keynote addresses, invited seminars, and media briefings. His research interests span several areas in parallel and distributed computing, as well as complex systems. Dr. Zomaya is an elected fellow of the Australian Academy of Science, the Royal Society of New South Wales, and an Elected Foreign Member of Academia Europaea. He is a decorated scholar with numerous accolades, including the Fellowship of the American Association for the Advancement of Science and the Institution of Engineering and Technology, U.K. He is a recipient of the IEEE Technical Committee on Parallel Processing Outstanding Service Award (2011), the IEEE Technical Committee on Scalable Computing Medal for Excellence in Scalable Computing (2011), the IEEE Computer Society Technical Achievement Award (2014), the ACM SIGSIM Reginald A. Fessenden Award (2017), the New South Wales Premier’s Prize of Excellence in Engineering and Information and Communications Technology (2019), the Research Innovation Award from the IEEE Technical Committee on Cloud Computing (2021), the Technical Achievement and Recognition Award, IEEE Communications Society’s IoT, Ad Hoc, and Sensor Networks Technical Committee (2022), and the Distinguished Technical Achievement Award, IEEE Communications Society’s Technical Committee on Big Data (2024). Dr. Zomaya is a Clarivate 2022\&2023 Highly Cited Researcher. He is the past Editor-in-Chief of the ACM Computing Surveys, IEEE TRANSACTIONS ON COMPUTERS, and the IEEE TRANSACTIONS ON SUSTAINABLE COMPUTING.
\end{IEEEbiography}

\vfill

\end{document}